\documentclass[10pt,pdflatex]{article}
\usepackage{wrapfig,cancel}
\usepackage{amsmath}%
\usepackage{amsfonts}%
\usepackage{amssymb}%
\usepackage{graphicx}
\usepackage[usenames]{color}
\numberwithin{equation}{section}
\newcommand \mathbox[1]{
	\vbox{\hrule
		\hbox{\vrule\kern8pt
			\vbox{\kern8pt 
				\hbox{
					$\displaystyle #1$
				}\kern8pt
			}\kern8pt\vrule
		}\hrule
	}
}
\def\omit#1{{}}
\newcommand{\tops}[2]{{#1}}
\definecolor{darkblue}{rgb}{0,0,.8}
	
\definecolor{red}{rgb}{1,0,0}
	
\def\BK#1{ \left\langle \rule{0pt}{10pt} #1 \right\rangle}

\newcommand{\Res}{\operatorname{Res}\displaylimits}

\def\e#1{\text{e}^{#1}}

\def\pd{\partial}
\def\tr#1{\text{tr}\left(#1\right)}
\def\Tr#1#2{{\text{tr}}_{{#1}}\left(#2\right)}
\def\det#1{\text{det}\left(#1\right)}
\def\d#1{\text{d}{#1}}

\def\OO#1{{\cal{O}}\left(#1\right)}

\def \id {\mathbb{I}}

\def \A {{\cal{A}}}
\def \C {\mathbb{C}}
\def \H {{\cal H}}
\def \R {\mathbb{R}}

\def \w {{\omega}}

\def \ep {\epsilon}
\def \and {\text{and }}

\def \ri{\right}
\def\ZC{{\cal Z}_C}

\def\slint{\setminus \hspace{-9.5pt} \int }

\def\ds{\displaystyle}

\def\H{{H\cdot}}
\def\F{{\cal F}}
\def\tF{{\tilde{\cal F}}}
\newtheorem{lemma}{Lemma}[section]
\newtheorem{remark}{Remark}[section]
\newtheorem{proposition}{Proposition}[section]
\def \be{\begin{equation}}
\def \ee{\end{equation}}
\def \bea{\begin{eqnarray}}
\def \eea{\end{eqnarray}}
\def\A{{\cal A}}
\def\B{{\cal B}}
\def\x{{x}}
\def\pa{\partial}
\def\calC{{\mathcal{C}}}
\def\calM{{\mathcal{M}}}

\def\K{{\bf K}}
\def\J{{\bf J}}
\newcommand{\Remark}[1]{{\noindent {\footnotesize {\bf Remark:} #1}}}
\newtheorem{theorem}{Theorem}[section]
\begin{document}
\baselineskip 15pt plus 1pt minus 1pt

\vspace{0.2cm}
\begin{center}
\begin{Large}
\fontfamily{cmss}
\fontsize{17pt}{27pt}
\selectfont
\textbf{Topological Expansion for the Cauchy two-Matrix-Model}
\end{Large}\\
\bigskip
\begin{large} {
M.
Bertola}$^{\ddagger,\sharp}$\footnote{Work supported in part by the Natural
    Sciences and Engineering Research Council of Canada
(NSERC).}\footnote{bertola@crm.umontreal.ca}, A. Prats Ferrer$^{\sharp}$\footnote{pratsferrer@crm.umontreal.ca}
\end{large}
\\
\bigskip
\begin{small}
$^{\ddagger}$ {\em Department of Mathematics and
Statistics, Concordia University\\ 1455 de Maisonneuve W., Montr\'eal, Qu\'ebec,
Canada H3G 1M8} \\ $^{\sharp}$ {\em Centre de recherches math\'ematiques, Universit\'e\ de Montr\'eal\\ 2920 Chemin de la tour, Montr\'eal, Qu\'ebec, Canada H3T 1J4.}
\end{small}

\bigskip
{\bf Abstract}
\end{center}
Recently a two-matrix-model with a new type of interaction \cite{BeGeSz-08.2} has been introduced and analyzed using bi-orthogonal polynomial techniques.
Here we present the complete $\frac{1}{N^2}$ expansion for the {\it formal} version of this model, following the spirit of \cite{EyOr-05.1,ChEy-06.1}, i.e. the full expansion for the non mixed resolvent correlators and for the free energies.

\vspace{0.7cm}
\tableofcontents
\section{Introduction}\label{Sec:Int}

The relationship between matrix models and  graph counting has attracted attention for a long time. Such a connection hinges  on the interpretation of Feynmann diagrams in the formal expansion as random lattices with or without matter (see \cite{diFGiZiJ-95.1,BrItPaZu-78.1,Da-85.1,KaKoMi-85.1} for small selection of important works). Very recently the full structure of this  expansion has been unveiled \cite{Ey-03.1,EyOr-05.1,ChEy-06.1,ChEyOr-06.1}. The construction relies on the definition of a formal matrix model that provides the generating function for counting of fat graphs embedded in Riemann surfaces, and the topological expansion is the organization of these by the Euler characteristics of the Riemann surface in which they are embedded. 
The first type of matrix models where this structure was completely revealed was the one (Hermitian) matrix model, which counts orientable graphs with no other decoration \cite{Ey-04.1,ChEy-06.1}.

In multi-matrix models usually only the polynomial interaction in the action is considered. Apart from some isolated works, most of the attention has been directed towards the so-called Harish--Chandra-Itzykson-Zuber type interaction which appears to be the only generically integrable case\footnote{In some sense it is the only Gaussian interaction possible.}. The first multi-matrix model for which the full topological expansion was revealed was the two (Hermitian) matrix model, and happened to formally share the same structure \cite{Ey-03.1,EyOr-05.1,ChEyOr-06.1}. Similarly, in \cite{EyPr-08.1} the same structure was found describing the full topological expansion of the chain of (Hermitian) matrices.

This apparent {\it universal} description relies  only upon the existence of an algebraic curve describing the leading order of the topological expansion. In \cite{EyOr-07.1} the same construction was developed for any (non-singular) algebraic curve without considering any underlying matrix model. The construction has been successfully applied in connection to certain non-algebraic curves and in relation with algebraic-geometric counting problems (see \cite{EyMaOr-07.1,EyOr-07.2,Ey-07.1} as an example).

Although the general case has already been worked out, the combinatorial interpretation is not clear for an arbitrary algebraic curve, except for some isolated particular cases. 

In this work we show how the same topological expansion construction works in a recently introduced two (Hermitian) matrix model called Cauchy two matrix model \cite{BeGeSz-08.2}.

In section \ref{Sec:M} the Cauchy matrix model is recalled. The eigenvalue representation is presented and then written back into the matrix representation in a much more convenient form.\\
In section \ref{Sec:LESC} the two master loop equations needed for the construction are presented. The main difference in the derivation (contained in App. \ref{App:CoV}) is that the domain of integration has a natural boundary since it extends only over the {\em positive definite} matrices. This yields some poles in the structural master loop equations.\\
In section \ref{Sec:3MLESC} the algebraic curve (first step of the construction) is found as the leading order contribution from one of the master loop equations. This will provide the leading order solution for the one point resolvent. Everything is expressed in terms of the algebraic curve.\\
In section \ref{Sec:2MLE} the other master loop equation's topological expansion is analyzed. This leads to a recursion equation between the different orders in the topological expansion of $1$ and $2$ point resolvents. The recursion equation is solved and the solution for general $k$-point resolvents is presented.\\
In section \ref{Sec:TEFE} the topological expansion of the free energy (thus that of the partition function itself) is found by inverting the loop insertion operator on the one point correlator. All the terms except the order zero and the order one can be found in that way.\\
In the appendices we have detailed some of the calculations that do not bring great inside into the problem or that are fundamentally equivalent to similar calculations in other works.\\
In \ref{App:AC} we give the definitions of the objects related to the algebraic curve that we use, and give some relations and properties.\\
In \ref{App:F01} we compute the order one term in the topological expansion of the free energy that could not be found by inverting the loop insertion operator. The computation is relatively  standard and it is included for convenience of the reader.\\

\section{The Cauchy Matrix Model}\label{Sec:M}

Consider the following matrix integral
\begin{equation}
	\ZC=\int_{M_i=M_i^\dag>0}\hspace{-15pt}\d{M_1} \d{M_2} \frac{\e{-\frac{N}{T}
	\tr{V_1(M_1)+V_2(M_2)}}}{\det{M_1+M_2}^N}=\e{-\frac{N^2}{T^2} \F}
	\label{eq:CMM}
\end{equation}
where the domain of integration (as indicated) is the ensemble of $N\times N$ Hermitian {\em positive definite} matrices. 
We will view the matrices $M_j$ as random variables with the probability measure 
\begin{equation}
{\rm d}P(M_1,M_1):=\frac 1{\ZC }\d{M_1} \d{M_2} \frac{\e{-\frac{N}{T}
	\tr{V_1(M_1)+V_2(M_2)}}}{\det{M_1+M_2}^N}
\end{equation}
The {\it potentials} $V_i$ 
parametrize the measure associated to each matrix, and must satisfy a suitable growth condition at infinity  for the integral to be convergent; as typical we will require $\displaystyle V_i(x)/\ln (x) \mathop{\longrightarrow}_{x\to+\infty}+\infty$. 

We will also assume that they diverge at the origin of the spectra with at least logarithmic growth. The simplest class of such potentials, to which we restrict in this paper, is the class of ``polynomials + logarithm''
\begin{equation}
	V_i(x)=-t_{-1}^{(i)}\ln{x} + \sum_{k=1}^{d_k} t_k^{(i)} x^k\ ,\ \ t_d^{(i)} >0, t_{-1}^{(i)} >0\ ,\ \ i=1,2
\end{equation} 
although later we will allow (formal) variations with respect to the infinite number of parameters $t_k^{(i)}$.
The log term is included to avoid the eigenvalues approaching the hard edge of the spectrum of positive matrices at $x=0$. 
The expression 
$\F$ is called the free energy of the matrix model and we call $T$ the total charge.

Thanks to the Harnad-Orlov formula \cite{HaOr-06.1} and following \cite{BeGeSz-08.2} we can write \eqref{eq:CMM} in eigenvalue representation 
\begin{equation}
	\ZC=\int_{\R_+^N} \left(\prod_{i=1}^N\d{x_i} \d{y_i}\right)\Delta^2(x_i)\Delta^2(y_i) 
	\frac{\e{-\frac{N}{T}\sum_{i=1}^N \left(V_1(x_i)+V_2(y_i)\right)}}
	{\prod_{i,j=1}^N(x_i+y_j)}
	\label{eq:evCMM}
\end{equation}
where $x_i$ (resp. $y_i$) are the (positive) eigenvalues of $M_1$ (resp. $M_2$), and $\Delta(a_i)=\det{a_i^{j-1}}_{1\le i,j\le N}=\prod_{i<j}^N(a_i-a_j)$ is the Vandermonde determinants.

Note that we can easily write \eqref{eq:evCMM} back into matrix representation with a different (but equivalent as long as we look at non-mixed correlators) interaction, namely
\begin{equation}
	\ZC=\int_{M_i>0}\hspace{-15pt}\d{M_1} \d{M_2} 
	\frac{\e{-\frac{N}{T}\tr{V_1(M_1)+V_2(M_2)}}}
	{\det{M_1\otimes\id+\id\otimes M_2}}
	\label{eq:tpCMM}
\end{equation}
where now the determinant is acting on a $N^2\times N^2$ size matrix, and $\id$ represents the $N\times N$ identity matrix.
Note that the $N$ exponent in the determinant from \eqref{eq:CMM} has disappeared.

We will consider here the formal perturbative expansion of the matrix integral around one local minima.

The goal of the following sections is to analyze the loop equations for this matrix model and solve them in the spirit of \cite{EyOr-05.1}. 
For this we have to restrict ourselves to the formal version of the model, i.e. by choosing one of the many local extrema of the potentials and perform a perturbation of the action around it. This is in fact equivalent to the {\it fixed filling fraction} condition that amounts to specify and fix the amount of eigenvalues that lie close to a given extrema of the potential. 
Indeed a local extremum for the matrix action can be characterized by the number $N_j^{(k)}$ of eigenvalues of the matrix $M_k$ ($k=1,2$) that lie close to a given local extremum $j$ for the eigenvalue action. This numbers (after normalizing them) are called filling fractions.
All this considerations assure the existence of the topological expansion that we are trying to find.
Equation \eqref{eq:tpCMM} is the most convenient form of $\ZC$ for this type of analysis.

\section{Loop equations and spectral curve}\label{Sec:LESC}

This matrix model is somehow peculiar because instead of one master loop equation we will need two of them. One of them will provide the large $N$ algebraic curve that underlie all our calculation. The other contains, in its large $N$ expansion, the recurrence equations for the algebro-geometric objects that are naturally associated to the algebraic curve and that solve the model.
Since the derivation of the loop equations is a somehow standard technique, the ones we use in this paper are derived in appendix \ref{App:CoV}. In this section we only define the objects appearing in these loop equations and write down the equations before analyzing and extracting all information we need from them. The interested reader can find more details on their derivation in appendix \ref{App:CoV}.

\subsection{Definitions}\label{sSec:Def}

 Let us first define the following scalar random variables depending on the spectral parameter $x$,
\begin{equation}
\begin{split}
	W_1(x)&=\frac{T}{N}\tr{\frac{1}{x-M_1}}\\
	P_1(x)&=\frac{T}{N}\tr{\frac{V_1^\prime(x)-V_1^\prime(M_1)}{x-M_1}}\\
	Q_1(\x)=&{\frac{T^2}{N^2}}\tr{\frac{V_1^\prime(\x)-V_1^\prime(M_1)}{\x-M_1}
	\Tr{2}{\frac{1}{M_1\otimes\id+\id\otimes M_2}}}
\end{split}	
\end{equation}
and of $M_2$ and the same spectral parameter $x$
\begin{equation}
\begin{split}
	{W}_2(x)&=-\frac{T}{N}\tr{\frac{1}{x+M_2}}\\
	{P}_2(x)&=-\frac{T}{N}\tr{\frac{V_2^\prime(-x)-V_2^\prime(M_2)}{x+M_2}}\\
	{Q}_2(-\x)=&{\frac{T^2}{N^2}}\tr{\frac{V_2^\prime(\x)-V_2^\prime(M_1)}{\x-M_1}
	\Tr{1}{\frac{1}{M_1\otimes\id+\id\otimes M_2}}}.
\end{split}	
\end{equation}
The notation $\Tr{i}{\dots}$ stands for the {\bf partial trace} on the corresponding factor of the tensor-product space.
The functions $W_i(x)$ are called resolvent and will be considered only as formal expansions in $x^{-1}$, i.e. as a generating function for traces of powers of $M_i$. The functions $P_i$ and $Q_i$ are meromorphic functions of $x$ with a pole only at $x=0$.

Finally we introduce the following $x$--independent scalar random variables
\begin{equation}
\begin{split}
	I_1&=\frac{T}{N}\tr{V_1^\prime(M_1)}+\frac{T^2}{N^2}\Tr{1,2}{\frac{1}{M_1\otimes\id +\id\otimes M_2}}\\
	{I}_2&=-\left(\frac{T}{N}\tr{V_2^\prime(M_2)}
	+{\frac{T^2}{N^2}}\Tr{1,2}{\frac{1}{M_1\otimes\id +\id\otimes M_2}}\right).
\end{split}
\end{equation}
The dependence on $M_i$ of $I_{1,2}$ will be understood. We introduce the following notation for the average
\begin{equation}
	\BK{(\cdots)}=\frac{1}{\ZC}\int_{M_i>0}\hspace{-15pt}\d{M_1} \d{M_2} \,(\cdots)\, 
	\frac{\e{-\frac{N}{T}\tr{V_1(M_1)+V_2(M_2)}}}{\det{M_1\otimes\id+\id\otimes M_2}}
\end{equation}
Finally we define the following auxiliary functions for later convenience in writing the loop equations
\begin{equation}
\begin{split}
	Y_1(\x)&=U^\prime_1(\x)-W_1(\x)\,,\quad U^\prime_1(\x)=\frac{2V_1^\prime(\x)-V_2^\prime(-\x)}{3}\\
	{Y}_2(\x)&=U^\prime_2(\x)-{W}_2(\x)\,,\quad U^\prime_2(\x)=\frac{-V_1^\prime(\x)+2V_2^\prime(-\x)}{3}
\end{split}
\end{equation}
They differ from the corresponding random variables only in a deterministic shift that depends explicitly on the potentials.
Although it may seem redundant we also need the following definitions
\begin{equation}
\begin{split}
	W_0(\x)&:=-W_1(\x)-{W}_2(\x)\\
	Y_0(\x)&:=-Y_1(\x)-{Y}_2(\x)=U^\prime_0(\x)-W_0(\x)
\end{split}
\end{equation}

\subsection{The Master Loop Equations}

The first of the two master loop equations is quadratic in both $Y_1$ and $Y_2$ and has the form
\begin{equation}
	\BK{\left(Y_1(\x)\right)^2}+\BK{\left({Y}_2(\x)\right)^2}+\BK{Y_1(\x){Y}_2(\x)}
	=\BK{R(\x)}-\frac{1}{\x}\BK{I_1+{I}_2}:=\BK{\hat{R}(\x)}
	\label{eq:YQMLE}
\end{equation}
where
\begin{equation}
\begin{split}
	R(\x)&=\frac{1}{3}(V_1^\prime(\x)^2+V_2^\prime(-\x)^2-V_1^\prime(\x)V_2^\prime(-\x))-P_1(\x)-{P}_2(\x)\\
	      &=(U^\prime_1(\x))^2+(U^\prime_2(\x))^2+U^\prime_1(\x)U^\prime_2(\x)-P_1(\x)-{P}_2(\x)
\end{split}
	\label{eq:RDef}
\end{equation}
is a meromorphic  function with poles  of order at most $2$  only at $x=0$.

The second of the two master loop equations is a cubic equation that both $Y_1$, $Y_2$ and $Y_3$ satisfy identically  
\begin{equation}
\begin{split}
	\BK{Y_k(\x)^3}-\BK{\hat{R}(\x)Y_k(\x)}
	-{\frac{T^2}{N^2}}\left(\frac{1}{2}\frac{\d{}^2}{\d{\x}^2}+\frac{1}{\x}\frac{\d{}}{\d{\x}}\right)
	\BK{W_k(\x)}&=\BK{D(\x)}\,,\quad \text{for $k=1,2$}\\
\end{split}
	\label{eq:YCMLE.a}
\end{equation}
where 
\begin{equation}
\begin{split}
	\BK{D(\x)}&=-U^\prime_0(\x)U^\prime_1(\x)U^\prime_2(\x)-U^\prime_1(\x)\BK{{P}_2(\x)}-U^\prime_2(\x)\BK{P_1(\x)}
	-\frac{1}{\x}\left(U^\prime_1(\x)\BK{{I}_2}+U^\prime_2\BK{I_1}\right)\\
	&\qquad +\BK{Q_1(\x)+{Q}_2(\x)}+\BK{W_1(\x)^2{W}_2(\x)}+\BK{W_1(\x){W}_2(\x)^2}+\BK{S_1(\x)}+\BK{{S}_2(\x)}
\end{split}
	\label{eq:DDef}
\end{equation}
is a meromorphic function with pole at $x=0$.
The last two terms are defined in the appendix \ref{App:CoV}. The last four terms contribute only to the residue at $x=0$.

\section{The cubic master loop equation and the spectral curve}\label{Sec:3MLESC}

Before going any further let us rewrite equation \eqref{eq:YCMLE.a} in a suitable form for latter use.
As usual with matrix models loop equations we rewrite everything in terms of connected correlators.

Let $A$ be any scalar random variable obtained as trace of a polynomial of $M_1$.
We define the connected part of a two-trace correlator as
\begin{equation}
	\BK{A(\xi)W_1(\xi^\prime)}_c={\frac{N^2}{T^2}}\left(\BK{A(\xi)W_1(\xi)}-\BK{A(\xi)}\BK{W_1(\xi)}\right)
	=\frac{\pd{}}{\pd{}V_1(\xi^\prime)}\BK{A(\xi)} - 
	\BK{\frac{\pd{}A(\xi)}{\pd{}V_1(\xi^\prime)}}
	\label{eq:ConCor}
\end{equation}
with the definition of the {\em vertex operator}
\begin{equation}
	\frac{\pd{}}{\pd{}V_1(\xi)}=-\sum \frac{1}{\xi^{i+1}}\frac{\pd{}}{\pd{}{t_i}}\,,\quad 
	V_1(\xi)=\sum t_i \xi^i.
	\label{eq:ILO}
\end{equation}
This formula is valid even when $A(\xi)$ depends on $V_1$.

\begin{remark}
Note that the last equality reduces to the usual definition whenever $A(\xi)$ is independent of the moduli in $V_1$. When $A$ depends explicitly on $V_1$ we need to add the extra term in the last equality of \eqref{eq:ConCor} to compensate for this dependence.
\end{remark}

In particular we are interested in the following relations
\begin{equation}
\begin{split}
	\BK{\left(Y_k(\xi)\right)^2}&=\BK{Y_k(\xi)}^2+{\frac{T^2}{N^2}}\BK{W_{k}(\xi)W_{k}(\xi)}_c\\
	\BK{Y_k(\xi){Y}_j(\xi)}&=\BK{Y_k(\xi)}\BK{{Y}_j(\xi)}+\frac{T^2}{N^2}\BK{W_k(\xi)W_k(\xi)}_c\\
	\BK{\hat{R}(\xi)W_k(\xi)}&=
	\BK{\hat{R}(\xi)}\BK{W_k(\xi)}+\frac{T^2}{N^2}\BK{\overline{R}(\xi)W_k(\xi)}_c
\end{split}
	\label{eq:CCor.a}
\end{equation}
where we have defined
\begin{equation}
	\overline{R}(\xi)=-P_1(\xi)-{P}_2(\xi)-\frac{1}{\xi}(I_1+{I}_2)
	\label{eq:CCaDef.a}
\end{equation}
Similarly we obtain
\begin{equation}
	\BK{\left(Y_k(\xi)\right)^3}=\BK{Y_k(\xi)}^3+\frac{T^2}{N^2}3\BK{Y_k(\xi)}\BK{W_k(\xi)W_k(\xi)}_c
	-\frac{T^4}{N^4}\BK{W_k(\xi)W_k(\xi)W_k(\xi)}_c.
	\label{eq:CCaDef.b}
\end{equation}
Putting all together we have that \eqref{eq:YCMLE.a} becomes
\begin{equation}
\begin{split}
	\BK{Y_k(\xi)}^3-\BK{\hat{R}(\xi)}\BK{Y_k(\xi)}=&\BK{D(\xi)}
	+\frac{T^4}{N^4}\BK{W_k(\xi)^3}_c\\
	+&\frac{T^2}{N^2}\Bigg(\left(\frac{1}{2}\frac{\d{}^2}{\d{\xi}^2}+\frac{1}{\xi}\frac{\d{}}{\d{\xi}}\right)\BK{W_k(\xi)}
	-\BK{\overline{R}(\xi)W_k(\xi)}_c
	-3\BK{Y_k(\xi)}\BK{(W_k(\xi))^2}_c\Bigg)
\end{split}
	\label{eq:YCMLE.b}
\end{equation}

\subsection{The large \tops{$N$}{N} limit of the cubic master loop equation}

We now postulate  a $T^2/N^2$ expansion of the form\footnote{Indeed, by definition the $1/N^2$ expansion of $\BK{Y_k(\xi)}$ is the same as that of $W_k(\xi)$ except for the leading term.}
\begin{equation}
\begin{split}
	\BK{Y_k(\xi)}&=y_k(\xi)-\sum_{h=1}^{\infty} \left(\frac{T}{N}\right)^{2h} W^{(h)}_k(\xi)\\
	\BK{\hat{R}(\xi)}&=\hat{R}^{(0)}(\xi)+\sum_{h=1}^{\infty} \left(\frac{T}{N}\right)^{2h} \overline{R}^{(h)}(\xi)\\
	\BK{D(\xi)}&=D^{(0)}(\xi)+\sum_{h=1}^{\infty} \left(\frac{T}{N}\right)^{2h} D^{(h)}(\xi)
\end{split}
	\label{eq:LN.a}
\end{equation}
and take the leading term of \ref{eq:YCMLE.b}
\begin{equation}
	y_k(\xi)^3-\hat{R}^{(0)}(\xi)y_k(\xi)=D^{(0)}(\xi),
	\label{eq:CLN.a}
\end{equation}
The large $N$ limit of the cubic loop equation gives us the spectral curve
\begin{equation}
\begin{split}
	E(\x,y)=y^3-\hat{R}^{(0)}(\x)y-D^{(0)}(\x)=0
\end{split}
	\label{eq:SC}
\end{equation}
whose solutions are encoded in the large $N$ limits of $\BK{Y_k(x)}$
\begin{equation}
y=y_k(\x):= \lim_{N\to\infty}\BK{Y_k(x)}\ ,\qquad k=0,1,2.
\end{equation}
The relation $\sum_{k=0}^2 y_k(\x)=0$ (which is obvious from the definition of $Y_0(\x)$) is the reason of the absence of the quadratic term in the cubic spectral curve (\ref{eq:SC}).

The meaning of  Ansatz (\ref{eq:LN.a}) --as usual in these computations-- is that it is {\em consistent} with the formal manipulations of the loop equations, which is our perspective in this work. The expansion for the large $N$ expectations in the convergent model typically will not be of the advocated form, in particular if the asymptotic spectral curve (\ref{eq:SC})  has a genus other than zero. The {\em fixed filling fraction} condition placed us from the beginning away from the convergent model (except for the genus zero case also known as one cut case).
\subsection{The spectral curve}\label{ssSec:EC}

We now analyze the basic structure of the spectral curve $E(\x,y)$ that we denote by $\Sigma$.
In terms of the $\x$ variable $\Sigma$ is a three-sheeted covering of $\C$. Each sheet is parametrized by one of the solutions $y_k(\x)$ of the cubic equation \eqref{eq:CLN.a}. It was shown in \cite{BeGeSz-08.2, BertolaBalogh} (in the context of the {\em convergent matrix integral}) that the sheets $1$ and $0$ are glued along one or more cuts contained in  $\R_+$ and sheets $2$ and $0$ along cuts contained in $\R_-$. 

We will think of the three functions $y_k(\x)$ as one single valued function $y(p)$ on $\Sigma$ where $p\in\Sigma$. 
Similarly we think of $\x$ as a function $\x:\Sigma \to \mathbb P^1$  so that $E(\x(p),y(p))=0$ for all $p\in \Sigma$. 
Let us call $\chi_k$ the sheet defined by $y_k(\x)$, i.e.
\begin{equation}
	\chi_k=\{p\in\Sigma \text{ such that } y(p)=y_k(\x(p)) \}
	\label{eq:defXk}
\end{equation}
In general we will always view all the functions previously introduced as single--valued on the spectral curve, for example, the ones collecting $W_k(x)$ and $U^\prime_k(x)$
\begin{equation}
	\begin{split}
		W(p)&=\BK{W_k(x(p))} \qquad \text{whenever } p\in\chi_k\\
		U^\prime(p)&=U^\prime_k(x(p)) \qquad \text{whenever } p\in\chi_k
	\end{split}
	\label{eq:defWU}
\end{equation}

\begin{wrapfigure}{r}{0.6\textwidth}
\resizebox{0.6\textwidth}{!}{\input{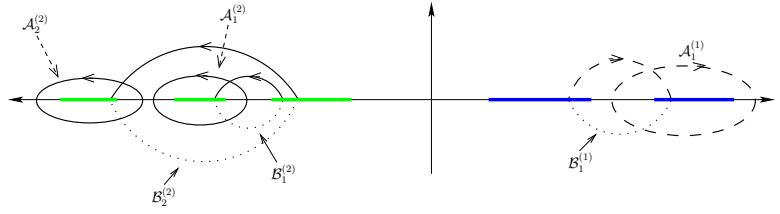}}
\caption{The choice of homology basis. The dashed contours run on $\chi_1$, the solid contours on $\chi_2$ and the dotted contours on $\chi_0$.}
\label{fighomo}
\end{wrapfigure}

Since the projection $\x:\Sigma \to \mathbb P^1$ is three-sheeted, there are three different points $p^{(i)}$ ($i=1,2,3$), such that $\x(p^{(i)})=\x(p^{(j)})$ for all $i$ and $j$. 
Whenever we write superscripts to the points on the curve, we will be referring to points on the same $\x$-projection. There are three different points above $\x=\infty$ for $y(p)$, one in each sheet. We will call them $\infty_k$, and in a neighbourhood of $\infty_k$ we have the behavior $y(x)\sim U^\prime_k(x(p))$.
Other important points are the endpoints of the cuts. These are solutions to the equation $\d x(p)=0$ and are the only points belonging more than one sheet. We assume that they are simple zeros of $\d x(p)$ which means that these points belong to exactly two sheets. 

 We denote the endpoints of the cuts by  $a_i^{(k)}$ ($k=1,2$, $i=0,\dotsc,2\bar{d}_k-1$, with $\bar{d}_{k}\leq{d}_{k}$) where the superscript indicates whether they belong to $\chi_{1}\cap\chi_0$ or $\chi_{2}\cap\chi_0$.  Assume that $|a_0^{(k)}|<|a_1^{(k)}|<\dotsm<|a_{2\bar{d}_k-1}^{(k)}|$, then the cuts are the intervals $A_i^{(k)}=[a_{2i}^{(k)},a_{2i+1}^{(k)}]$ for $i=0,\dotsc,\bar{d}_k-1$.
We will call $\alpha_i$ a generic branch point $a_j^{(k)}$ when there is no need to specify to which family it belongs.

We must also introduce a canonical choice of homology basis for $\Sigma$. Not all the cuts of each species are independent since the sum of them all is reducible to a point. There are $\bar{d}_k-1$ independent cuts of the $\A^{(k)}$ species\footnote{Remember that $d_k$ is the degree of the polynomial potential $V_k^\prime(x)$, and that $\bar{d}_k\leq d_k$.}, thus the maximal genus for $\Sigma$ is $\bar{d}_1+\bar{d}_2-2$.
We define the cycles $\A_i^{(1)}$ on $\chi_1$ encircling only the cut $A_i^{(1)}$ for $i=0,\dotsc,\bar{d}_1-1$ and similarly the cycles $\A_i^{(2)}$ on $\chi_2$ encircling only the cut $A_i^{(2)}$ for $i=0,\dotsc,\bar{d}_2-1$, and choose as independents all except $\A_0^{(k)}=-\sum_{j=1}^{d_k-1}\A_j^{(k)}$ ($k=1,2$). The conjugated cycles $\B_i^{(k)}$ go through the $0$th and the $k$th cuts on $\chi_k$ visiting both $\chi_k$ and $\chi_0$.
It is easy to verify that those cycles form a canonical basis (see Fig. \ref{fighomo}).
\subsubsection{ Moduli of the model.}

The data that defines the model is contained in 
\begin{itemize}
\item the values of the potential parameters $t_j^{(k)}$;
\item  the total charge $T$;
\item and the filling fractions $\epsilon_i^{(k)}$ defined below in eq. (\ref{eq:FFFC}).
\end{itemize}
 We denote a generic setting of these parameters as $\calM=\{t_j^{(k)},T,\epsilon_i^{(k)}\}$.
We will be interested in computing variations (derivatives) with respect to these parameters of different functions of them. In particular, for the potential parameters, we introduce a {\it generating operator} of such variations.
From the matrix model point of view the so called loop insertion operators are an important object. They define the one point resolvent when applied to the free energy, and define the connected $n+1$-point resolvent when applied to the connected $n$-point resolvent. In our model we have two of them: one for each matrix
\begin{equation}
	\begin{split}
		\frac{\pd{}}{\pd{}V_1(x)}&=-\sum_k \frac{1}{x^{1+k}}\frac{\pd{}}{\pd{}t_k^{(1)}}\\
		\frac{\pd{}}{\pd{}V_2(-x)}&=-\sum_k \frac{1}{(-x)^{1+k}}\frac{\pd{}}{\pd{}t_k^{(2)}}.
	\end{split}
	\label{eq:LIO}
\end{equation}
With these two operators we will define the combined loop insertion operator in $\Sigma$
\begin{equation}
	\frac{\pd{}}{\pd{}V(p)}=\left\{
	\begin{array}{ll}
		\frac{\pd{}}{\pd{}V_1(x)} & \text{whenever } p\in\chi_1\\
		-\frac{\pd{}}{\pd{}V_1(x)}-\frac{\pd{}}{\pd{}V_2(-x)} & \text{whenever } p\in\chi_0\\
		\frac{\pd{}}{\pd{}V_2(-x)} & \text{whenever } p\in\chi_2
	\end{array}  
	\right.
	\label{eq:LIOP}
\end{equation}
We remark that from the above definitions it follows by a direct computation that 
\begin{eqnarray}
&& -\frac{\pd}{\pd V(p)} \F = W(p)\cr
&& -\frac{\pd}{\pd V(p)}U^\prime(p^\prime) = \frac{2}{3}\frac{1}{(x(p)-x(p^\prime))^2}
\end{eqnarray}

\subsubsection{Fundamental differential on the algebraic curve}

The curve $\Sigma$ can be parametrized equally well by $\calM$ or by the meromorphic functions $y(p)$ and $x(p)$ (defined on some union of open sets in $\C$). In fact we will see that the differential $y(p)\d x(p)$ on $\Sigma$ contains the information in $\calM$.
Indeed,we can extract $t_k^{(j)}$ for $k>0$ from the behavior of $y\d x$ around\footnote{The behavior around the third infinity point $p=\infty_0$ is not independent due to the relation $y_1(x)+y_2(x)+y_3(x)=0$.} $\infty_{1,2}$
\begin{equation}
	\begin{split}
		\Res_{p\to\infty_1} \frac{y(p)\d x(p)}{x^{k}} &= \frac{2t_k^{(1)}-(-1)^k t_k^{(2)}}{3}\\
		\Res_{p\to\infty_2} \frac{y(p)\d x(p)}{x^{k}} &= \frac{-t_k^{(1)}+(-1)^k 2t_k^{(2)}}{3}
	\end{split}
	\label{eq:PotPar}
\end{equation}
Also the filling fractions can be explicitly written as the $\A$ cycles of the fundamental differential $y(p)\d x(p)$
\begin{equation}
	\frac{1}{2\pi i}\oint_{\A_j^{(k)}} y(p) \d x(p) = 
	\epsilon_j^{(k)}=(-1)^{k}\frac{N_j^{(k)}(T+\eta^{(k)})}{N}.
	\label{eq:FFFC}
\end{equation}
The shifts $\eta^{(k)}$ in the total charge $T$ above are due to the residues $t_{-1}^{(k)},\ k=1,2$ at $0$ of the derivatives of the potentials and are related to those as follows,
\begin{equation}
	\begin{split}
		\eta^{(1)}&=\frac{2 t_{-1}^{(1)}+t_{-1}^{(2)}}{3}\\
		\eta^{(2)}&=\frac{2 t_{-1}^{(2)}+t_{-1}^{(1)}}{3}
	\end{split}
\end{equation}
The sign in \eqref{eq:FFFC} appears because of the reflection to the negative axis.
The filling fractions condition $\sum_{j=0}^{d_k-1} \epsilon_j^{(k)}=(-1)^k(T+\eta^{(k)})$ for both $k=1,2$ is another way of writting  $\A_0^{(k)}=-\sum_{j=1}^{d_k-1}\A_j^{(k)}$ and implies a relation for $\eta^{(k)}$ and the total charge $T$
\begin{equation}
	\begin{split}
		\Res_{p\to\infty_{0}} y(p)\d x(p)=&-\eta^{(1)}+\eta^{(2)}\\
		\Res_{p\to\infty_{1}} y(p)\d x(p)=&T+\eta^{(1)}\\
		\Res_{p\to\infty_{2}} y(p)\d x(p)=&-(T+\eta^{(2)})
	\end{split}
\end{equation}

\subsubsection{Multidifferentials on the algebraic curve}

In the next two sections we are going to compute the full topological expansion for the $n$-point resolvent correlator and the free energy, and in order to do so we need to define some functions and differentials on the curve.
Consider the $n$-point resolvent correlator
\begin{equation}
		W_{k_1,\dotsc,k_n}(\x(p_1),\dotsc,\x(p_n))=\BK{\prod_{i=1}^n W_{k_i}(x(p_i))}_c.\\
	\label{eq:wkdef}
\end{equation}
the construct a function on $\Sigma$ for every $n$
\begin{equation}
	W_n(p_1,\dotsc,p_n)=W_{k_1,\dotsc,k_n}(\x(p_1),\dotsc,\x(p_n))
	\hspace{20pt} 
	\begin{array}{c}
		\text{whenever $p_i\in\chi_{k_i}$}\\
		\text{for all $i=1,\dotsc,n$.}
	\end{array}
	\label{eq:hwkdif}
\end{equation}
and its topological expansion
\begin{equation}
	W_n(p_1,\dotsc,p_n)=\sum_{h=0}^\infty \left(\frac{T}{N}\right)^{2h} W_n^{(h)}(p_1,\dotsc,p_n).
\end{equation}
Finally, take $h$-order $n$-point resolvent correlator and construct the multi-differential
\begin{equation}
	\overline{\w}_n^{(h)}(p_1,\dotsc,p_n)=W_n^{(h)}(p_1,\dotsc,p_n)\prod_{i=1}^n \d \x(p_i)
	\label{eq:hwkdif.b}
\end{equation}
In particular we have the one-point differentials
\begin{equation}
	\overline{\w}_1^{(h)}(p)=W^{(h)}(p)\d x(p)
	\label{eq:hwdif}
\end{equation}
\Remark{Note that the notation changes and the subindex no longer indicates which sheet the point lies. Instead it indicates that it is a $n$-differential or $n$-variables function.}

Another way of writing it using the loop insertion operator is
\begin{equation}
	\overline{\w}_n^{(h)}(p_1,\dotsc,p_n)=\prod_{i=1}^n \d x(p_i)\prod_{i=1}^n \frac{\pd}{\pd V(p_i)} \F^{(h)}
	\label{eq:DIF6}
\end{equation}
where $\F^{(h)}$ is the $h$th term in the topological expansion of the free energy $\F$.

For latter use we define also
\begin{equation}
	\w_n^{(h)}(p_1,\dotsc,p_n)=\overline{\w}_n^{(h)}(p_1,\dotsc,p_n)
	+\delta_{n,2}\delta_{h,0}\frac{\d x(p_1)\d x(p_2)}{(x(p_1)-x(p_2))^2}
	\label{eq:wkbdef}
\end{equation}
which only modifies $\w_2^{(0)}(p)$ and leaves the rest untouched.

\subsection{Variations of the spectral curve with respect to the moduli}
We begin by studying how $y(p)\d x(p)$ changes under variation of the moduli. Since we have the constraint $E(x,y)=0$, $x(p), y(p)$ and $p$ are not independent, we have to keep one of them fixed while computing the variations. For convenience all variations are taken at $x(p)$ fixed. We will generically call this variations $\d\Omega$.
The variations will be written in terms of standard objects in algebraic geometry that we define in appendix \ref{App:AC}. The reader can find there all necessary information.

\subsubsection{Variation with respect to the filling fractions}
Not all filling fractions are independent due to the restrictions pointed out in the previous section.
As with $\A_j^{(k)}$ we consider as independent the $\epsilon_j^{(k)}$ for $j=1,\dotsc,d_k-1$. 
Clearly, since these ones are independent we have $\frac{\pd \epsilon_i^{(k)}}{\pd \epsilon_j^{(k^\prime)}}=\delta_{i,j}\delta_{k,k^\prime}$ for $i,j\not=0$.

Let us compute the variation of $y(p)\d x(p)$. We have that
\begin{equation}
	\frac{\pd}{\pd \epsilon_j^{(k)}} y(p)\d x(p) =\frac{\pd}{\pd \epsilon_j^{(k)}} (U^\prime(p)-W(p))\d x(p)= \OO{x(p)^{-2}}\d x(p)
	\label{eq:Ve1}
\end{equation}
when $x(p)\to\infty$. Indeed the potential $U^\prime(p)$ does not depend on $\epsilon$'s by hypothesis, and $W(p)$ gives us the behavior at $\infty$.
On the other side we can compute the $\A$ cycles of the above differential
\begin{equation}
	\begin{split}
		\frac{1}{2\pi i}\oint_{\A_j^{(k)}} \frac{\pd}{\pd \epsilon_{j^\prime}^{(k^\prime)}} y(p)\d x(p)&=
		\frac{\pd}{\pd \epsilon_{j^\prime}^{(k^\prime)}} \frac{1}{2\pi i}\oint_{\A_j^{(k)}} y(p)\d x(p)\\
		&=\frac{\pd \epsilon_j^{(k)}}{\pd \epsilon_{j^\prime}^{(k^\prime)}}=\delta_{j,j^\prime}\delta_{k,k^\prime}
	\end{split}
	\label{eq:Ve2}
\end{equation}
All together we have that $\frac{\pd}{\pd \epsilon_j^{(k)}} y(p)\d x(p)$ is a differential on $\Sigma$ with no pole and with normalizing condition \eqref{eq:Ve2}, which determine them completely to be a basis of normalized first type Abelian differentials defined in appendix \ref{App:AC} and that can be written in terms of the Bergmann Kernel $B(p,q)$ defined as well in appendix \ref{App:AC}
\begin{equation}
	\d\Omega_{\epsilon_j^{(k)}}\frac{\pd}{\pd \epsilon_j^{(k)}} y(p)\d x(p)= 2\pi i\d u_j^{(k)} = \oint_{\B_j^{(k)}} B(.,p) 
	\label{eq:Ve3}
\end{equation}

\subsubsection{Variation with respect to \tops{$\eta^{(k)}$}{etak} and the total charge \tops{$T$}{T}}

Imagine now that we change the total charge $T$, the variation of $y(p)\d x(p)$ will be
\begin{equation}
	\frac{\pd}{\pd T} y(p)\d x(p) = \left\{
	\begin{array}{ll}
		\OO{x(p)^{-2}}\d x(p) & \text{ $p\in\chi_0$}\\
		\left(\frac{-1}{x(p)}+\OO{x(p)^{-2}}\right)\d x(p) & \text{ $p\in\chi_1$}\\
		\left(\frac{1}{x(p)}+\OO{x(p)^{-2}}\right)\d x(p) & \text{ $p\in\chi_2$}
	\end{array}\right.
	\label{eq:VT1}
\end{equation}
Indeed by hypothesis the potentials do not depend on $T$, and the behavior at infinity of $W_k(x)$ gives us the above behavior.
Clearly, this differential has poles only at $\infty_{1,2}$ of order $1$ with residues $+1$, $-1$ respectively.
It is also easy to see that the $\A$ cycles of this differential are zero because the filling fractions do not depend on $T$ either. 
The conclusion is that
\begin{equation}
	\d\Omega_T(p)=\frac{\pd}{\pd T} y(p)\d x(p) = \d S_{\infty_1,\infty_2}(p) = \int_{\infty_2}^{\infty_1} B(. , p)
	\label{eq:VT2}
\end{equation}
which is a normalized third type Abelian differential (see appendix \ref{App:AC}).
Analogous calculations show that
\begin{equation}
	\begin{split}
		\d\Omega_{\eta^{(1)}}(p)
		&=\frac{\pd}{\pd \eta^{(1)}} y(p)\d x(p) = \d S_{\infty_1,\infty_0}(p) = \int_{\infty_0}^{\infty_1} B(. , p)\\
		\d\Omega_{\eta^{(2)}}(p)
		&=\frac{\pd}{\pd \eta^{(2)}} y(p)\d x(p) = \d S_{\infty_0,\infty_2}(p) = \int_{\infty_2}^{\infty_0} B(. , p)
	\end{split}
	\label{eq:VT3}
\end{equation}

\subsubsection{Variation with respect to the potentials}
Consider now the potentials. The variations with respect to the parameters of the potentials look like
\begin{equation}
	\begin{split}
		\d \Omega_{t_j^{(1)}} (p)\equiv
		\frac{\pd}{\pd t_j^{(1)}} y(p)\d x(p)&=\frac{\pd}{\pd t_j^{(1)}} (U^\prime(p)+T x(p)^{-1}+\OO{x(p)^{-2}}\d x(p)\\
		&=\left\{\begin{array}{ll}
			\frac{2}{3}j x(p)^{j-1}\d x(p)+\OO{x(p)^{-2}} &\text{for $p\in\chi_1$}\\
			-\frac{1}{3}j x(p)^{j-1}\d x(p)+\OO{x(p)^{-2}} &\text{for $p\not\in\chi_1$}
		\end{array}\right.\\
		\d \Omega_{t_j^{(2)}} (p)\equiv
		\frac{\pd}{\pd t_j^{(2)}} y(p)\d x(p)&=\frac{\pd}{\pd t_j^{(2)}} (U^\prime(p)-T x(p)^{-1}+\OO{x(p)^{-2}}\d x(p)\\
		&=(-1)^{j-1}\left\{\begin{array}{ll}
			\frac{2}{3}j x(p)^{j-1}\d x(p)+\OO{x(p)^{-2}} &\text{for $p\in\chi_2$}\\
			-\frac{1}{3}j x(p)^{j-1}\d x(p)+\OO{x(p)^{-2}} &\text{for $p\not\in\chi_2$}
		\end{array}\right.
	\end{split}
	\label{eq:DIF1}
\end{equation}
This behaviors at $\infty_k$ plus the normalizing conditions $\oint_{\A_i^{(k)}}\d\Omega_j^{(k^\prime)}=0$  determine completely $\d \Omega_j^{(k)}$ as a combination of second kind differentials (see appendix \ref{App:AC})
\begin{equation}
	\begin{split}
  \d \Omega_j^{(1)}(p) &= \frac{2}{3}\Res_{\infty_1} x(.)^j B(.,p)-\frac{1}{3}\Res_{\infty_2,\infty_0} x(.)^j B(.,p)\\
  \d \Omega_j^{(2)}(p) &= 
  (-1)^{j-1}\left(\frac{2}{3}\Res_{\infty_2} x(.)^j B(.,p)-\frac{1}{3}\Res_{\infty_1,\infty_0} x(.)^j B(.,p)\right)
	\end{split}
	\label{eq:VP1}
\end{equation}

\paragraph{Explicit vs implicit derivatives of} {$y\d x$.}
We claim that we can represent $y\d x$ 
\begin{equation}
	\begin{split}
	y \d x=& 2\pi i\sum_{k=1}^2\sum_{j=1}^{d_k-1}\epsilon_j^{(k)} \d u_i^{(k)}
		+T \d S_{\infty_1,\infty_2}(p)+\eta^{(1)} \d S_{\infty_1,\infty_0}+\eta^{(2)} \d S_{\infty_0,\infty_2}
		+\sum_{k=1}^2\sum_{j}t_j^{(k)}\d\Omega_j^{(k)}\\
		=& \sum_{t_a\in \mathcal{M}} t_a \d\Omega_{t_a}(p)
	\end{split}
	\label{eq:ydx}
\end{equation}
where $t_a$ is a collective notation for the whole set of parameters of our model $\mathcal{M}=\{t_j^{(k)},T,\epsilon_j^{(k)}
,\eta^{(k)}\}$. To see this, we  provisionally denote by $\eta$ the right-hand-side of (\ref{eq:ydx}); it is seen, from the definition of the various terms that $\eta -y \d x$ has no poles at $\infty_0,\infty_1,\infty_3$ and hence it is a holomorphic differential. On the other hand, from the definition of the filling fractions $\epsilon_j^{(k)}$ it also appears that it has vanishing $\mathcal A$--cycles. Standard theorems then guarantee then that $\eta -y\d x\equiv 0$. Therefore (\ref{eq:ydx}) yields an alternative representation of the fundamental differential $y\d x$. 

On the face of it  appears that $\partial_{t_a} y \d x$ is the same as the coefficient of the corresponding ``time'' $t_a$ in the r.h.s. of  (\ref{eq:ydx}), namely, the ``explicit'' derivative of the r.h.s. of (\ref{eq:ydx}). Note that this is not a trivial statement, since the spectral curve and all the differentials appearing as coefficients of the times in (\ref{eq:ydx}) {\em do depend} (implicitly) on the times as well.

Note that all variations $\d\Omega_{t_a}$ are expressible as an integral operator on the Bergmann kernel $B(p,q)$
\begin{equation}
	\begin{split}
		\d \Omega_{t_a} &= \frac{\pd}{\pd t_a} y(p)\d x(p) = \int_{\calC_{t_a}} B(p,\cdot) \Lambda_{t_a}(\cdot)\\
	\end{split}
	\label{eq:dOBK} 
\end{equation} 
where $\calC_{t_a}$ is the integration path associated with the variation with respect to $t_a$ and $\Lambda_{t_a} (p)$ is a meromorphic function on the curve.
For example, looking at the formulae above, we have $\calC_{t_j^{1}}=\frac{2}{3}\calC_{\infty_1}-\frac{1}{3}\left(\calC_{\infty_0}+\calC_{\infty_2}\right)$, and 
$\Lambda_{t_j^{(1)}}(p)= x(p)^{j}$, where $\calC_{\infty_k}$ is a small cycle around $\infty_k$.
We thus introduce the integral operators  associated to the various derivatives
\begin{equation}
\begin{array}{c|c|l}
\ds \frac {\partial}{\partial t_a}\square  & & \qquad \mathcal J_a(\square)\\[14pt]
\hline&\\
\ds \frac {\partial}{\partial t_{k}^{(1)}} \square & { \mapsto} &
\ds \left(\frac 23 \Res_{\infty_1}- \frac 1 3 \Res_{\infty_0, \infty_2}\right )  x^j \square\\[14pt]
\ds \frac {\partial}{\partial t_{k}^{(1)}} \square & { \mapsto} &
\ds \left(- \frac 23 \Res_{\infty_2}+ \frac 1 3 \Res_{\infty_0, \infty_1}\right)  (-x)^j \square\\[14pt]
\ds \frac {\partial }{\partial \epsilon_{j}^{(k)}}\square & {\mapsto}  & \ds \oint_{\mathcal B_j}\square\\[14pt]
\ds \frac \partial{\partial \eta^{(1)}} \square  & { \mapsto} & \ds\slint_{\infty_0}^{\infty_1} \square\\[14pt]
\ds \frac \partial{\partial \eta^{(2)}} \square  & { \mapsto} & \ds\slint_{\infty_2}^{\infty_0} \square\\[14pt]
\ds \frac \partial{\partial T} \square  & { \mapsto} & \ds \slint_{\infty_2}^{\infty_1} \square
\end{array}
\label{tableint}
\end{equation}
where the regularized integrals $\ds \slint$ are defined (if necessary) by subtraction of the singular part in the variable $x(p)$ (see \cite{Bert-06.1}) 
With this notation we have --compactly--
\begin{equation}
y(p) \d x(p)  = \sum_a t_a \mathcal J_a(B(\cdot, p))\label{ydxalt}
\end{equation}

\subsection{The two-resolvent correlation function}

We can now compute the action of the loop insertion operators on $y(p)\d x(p)$.
\begin{equation}
	\d x(p^\prime)\frac{\pd}{\pd V_1(x(p^\prime))} y(p)\d x(p)=
	-\sum_{j=0}^{\infty}\frac{\d \Omega_j^{(1)}(p) \d x(p^\prime)}{\x(p^\prime)^{j-1}}
	\label{eq:DIF2}
\end{equation}
Notice that a part from a constant factor in front depending on which sheet $p$ sits, the sum in the above formula produce a second order pole whenever $x(p)=x(p^\prime)$. This fact, together with the vanishing $\A$ periods condition, indicates that
\begin{equation}
	\begin{split}
	\d x(p^\prime)\frac{\pd}{\pd V_1(x)} y(p)\d x(p)&=
	-\frac{2}{3}B(q^{(1)},p)+\frac{1}{3}\left(B(q^{(0)},p)+B(q^{(2)},p)\right)\\
	&=-B(q^{(1)},p)+\frac{1}{3}\frac{\d x(q) \d x(p)}{(x(q)-x(p))^2}
	\end{split}
	\label{eq:DIF3}
\end{equation}
where $B(q,p)$ is the so called Bergmann kernel, and $q^{(k)}\in\chi_k$ satisfy $x(q^{(k)})=x(q)=x$.
In the last line we have used the relation
\begin{equation}
	\sum_{i=0}^2 B(q^{(i)},p)=\frac{\d x(q) \d x(p)}{(x(q)-x(p))^2}.
	\label{eq:BK2}
\end{equation} 
We obtain similar results by applying $\d x(q)\frac{\pd}{\pd V_2(-x(q))}$ or 
$-\d x(q)\left(\frac{\pd}{\pd V_1(x(q))}+\frac{\pd}{\pd V_2(-x(q))}\right)$.
All three operators combined in $\d x(q)\frac{\pd}{\pd V(q)}$ give
\begin{equation}
	\begin{split}
	\d x(q)\frac{\pd}{\pd V(q)} y(p)\d x(p)
	&=-B(q,p)+\frac{1}{3}\frac{\d x(q) \d x(p)}{(x(q)-x(p))^2}
	\end{split}
	\label{eq:DIF4}
\end{equation}
where $\pa/\pa V(x)$ was defined in (\ref{eq:LIOP}) and $U$ in (\ref{eq:defWU}).

It is now a simple calculation to find that
\begin{equation}
	\begin{split}
		\overline{\w}_2^{(0)}(q,p)\equiv\d x(q)\frac{\pd}{\pd V(q)} \w_1^{(0)}(p)&=
		\d x(q)\frac{\pd}{\pd V(q)} (U^\prime(p)-y(p))=\\
		&=B(q,p)-\frac{\d x(q) \d x(p)}{(x(q)-x(p))^2}\\
		&=-(B(q^+,p)+B(q^-,p))
	\end{split}
	\label{eq:DIF5}
\end{equation}
where $q^\pm$ are the two $q^{(i)}\not=q$.
Note that \eqref{eq:DIF5} implies $\w_2^{(0)}(p^\prime,p)=B(p^\prime,p)$. 
We can also write it as
\begin{equation}
	\begin{split}
		\overline{\w}_2^{(0)}(p^\prime,p^{(k)})=-\sum_{j\not=k}\w_2^{(0)}(p^\prime,p^{(j)})
	\end{split}
	\label{eq:sh.b}
\end{equation}
which will be used latter.

\section{The quadratic master loop equation}\label{Sec:2MLE}

Once we have extracted the algebraic curve from the cubic loop equation, only remains to find the tower of recursion relations deriving from the master--loop equations. The recursion relations for the terms of the topological expansion are contained in the topological expansion of the quadratic loop equation.

\subsection{Recursion relations and topological expansion}\label{sSec:LNQ}

Using \eqref{eq:YCMLE.0Def} we can rewrite equation \eqref{eq:YQMLE} as
\begin{equation}
	\sum_{k=0}^2\BK{Y_k(x)^2}=2\BK{\hat{R}(x)}.
	\label{eq:QMLE*}
\end{equation}
Indeed, introducing the definition \eqref{eq:YCMLE.0Def} in the LHS of 	\eqref{eq:QMLE*} we get twice the LHS of \eqref{eq:YQMLE}.
Write this in terms of connected components
\begin{equation}
\begin{split}
	\sum_{k=0}^2\BK{Y_k(\x)}^2=2\BK{\hat{R}(\x)}-\frac{1}{N^2}\sum_{k=0}^2\BK{W_{k}(\x)^2}_c
\end{split}
	\label{eq:LNQ}
\end{equation}
and extract the order $\frac{1}{N^{2h}}$ of the previous equation. We get the equation
\begin{equation}
	\begin{split}
		2\sum_{k=0}^2y(p^{(k)}) W_1^{(h)}(p^{(k)})
		=&-2\overline{R}^{(h)}(\x)+\sum_{k=0}^2\left(\sum_{m=1}^{h-1}W_1^{(m)}(p^{(k)})W_1^{(h-m)}(p^{(k)})
		+W_2^{(h-1)}(p^{(k)},p^{(k)})\right)
	\end{split}
	\label{eq:LNQh}
\end{equation}
for $h>0$.

Equation \eqref{eq:LNQh} is of the same form as the recursion equations solved in \cite{EyOr-05.1} and generalized in \cite{EyOr-07.1} to arbitrary algebraic curves. 
The final solution is written in the following way
\begin{equation}
	\begin{split}
		\w_1^{(h)}(q)&=-\sum_{\alpha}\Res_\alpha \frac{1}{2}\frac{\d S_{p,\bar{p}}(q)}{(y(p)-y(\bar{p}))\d x(p)}
		\left(\sum_{m=1}^{h-1}\w_1^{(m)}(p)\w_1^{(h-m)}(\bar{p})
		+\w_2^{(h-1)}(p,\bar{p})\right)\\
	\end{split}
\label{eq:Sol.a}
\end{equation} 
and can be generalized to 
\begin{equation}
	\begin{split}
		\w_{k+1}^{(h)}(q,p_{\K})
		&=-\sum_{\alpha}\Res_{p\to\alpha} 
		\frac{1}{2}\frac{\d S_{p,\bar{p}}(q)}{(y(p)-y(\bar{p}))\d x(p)}
		\left(\sum_{m=1}^{h-1}\sum_{\J\subset\K}\w_{j+1}^{(m)}(p,p_{\J})\w_{k-j+1}^{(h-m)}(\bar{p},p_{\K\backslash\J})
		+\w_{k+2}^{(h-1)}(p,\bar{p},p_{\K})\right)\\
	\end{split}
\label{eq:Sol.b}
\end{equation}
for higher correlators by applying $\frac{\pd}{\pd V(x)}$ to equation \ref{eq:Sol.a}.
The notation $p_\K$ stands for an array of variables $p_{i_1},p_{i_2},\dotsc,p_{i_k}$ where $\K=\{i_1,i_2,\dotsc,i_k\}$ is the set of subindices, $\J$ stands for subsets of subindices of $\K$, and $\K\backslash\J$ is the complementary subset of $\J$.
The derivation of these identities is contained in \cite{EyOr-05.1} and we refer ibidem for details.

\begin{remark} 
In those recursion relations, the initial data is contained in $y(p)\d x(p)$ and $\w_2^{(0)}(p,p^\prime)$. Note that the latter is defined by \eqref{eq:wkbdef} and \eqref{eq:DIF5} to be $\w_2^{(0)}(p^\prime,p)=B(p^\prime,p)$. It is not exactly equal to the 2-point resolvent correlator (remember the shift with respect to $\overline{\w}_2^{(0)}$). On the other side, the other differentials $\w_k^{(h)}$ do coincide with k-point resolvent correlators.
\end{remark}
\begin{remark}
It is also important to notice that by the differentials $\d S(q)$ in those formulas are defined through integral formulas, e.g.
\begin{equation}
	\d S_{p,\bar{p}}(q) = \int_{\bar{p}}^{p} B(q,r)
\end{equation} 
where the whole integration path must remain in a small neighbourhood of the corresponding $\alpha$, or be homologically equivalent.
\end{remark}

\subsection{Variations of \tops{$\w_k^{(h)}$}{the correlation forms} with respect to the moduli}

In \cite{EyOr-07.1} the authors proved (see section 5.1 and Theorem 5.1) the following relation
\begin{equation}
	\begin{split}
		\frac{\pd}{\pd t_a} \w_{k}^{(h)}(p_{\K})=\int_{\calC_{t_a}} \w_{k+1}^{(h)}(p_\K,q) \Lambda_{t_a}(q)
	\end{split}
\end{equation} 
for differentials satisfying \eqref{eq:Sol.b} and variations of the form \eqref{eq:dOBK}.
This equation is in fact essential in order to find the topological expansion of the free energy.
The proof is by induction supposing that it is satisfied for all $\w_k^{(h)}$ up to some $(k,h)$-level.
This important formula is telling us the way these correlators change when we change our model. This formula will be useful in the next section.

\section{Topological expansion of the free energy}\label{Sec:TEFE}

We will now compute the topological expansion of the free energy
\begin{equation}
	\F=\sum_{h=0}^\infty \left(\frac{T}{N}\right)^{2h-2}\F^{(h)}
	\label{eq:Fh} 
\end{equation} 

The free energy of the matrix model satisfy an homogeneity equation.
Suppose one scale the parameters of our model in the following way
\begin{equation}
	t_j^{(k)}\to \kappa t_j^{(k)} \,,\quad T \to \kappa T
\end{equation} 
the partition function of the model remains unchanged. Indeed the formal integral stays the same due to the fact that it only depends on $\frac{t_j^{(k)}}{T}$.
Notice that a side effect of this scaling is the corresponding scaling for the filling fractions $\epsilon_j^{(k)}\to \kappa \epsilon_j^{(k)}$. This means that changing $\kappa$ do not change $\F$
\begin{equation}
	\kappa\frac{\pd}{\pd \kappa} \F =0
\end{equation} 

Looking at \eqref{eq:Fh} we have that $\F^{(h)}$ should be homogeneous in $\kappa$ of order $2-2h$ 
\begin{equation}
	\kappa \frac{\pd}{\pd \kappa} \F^{(h)}= (2-2h)\F^{(h)}
	\label{eq:HE} 
\end{equation}
Equation \eqref{eq:HE} can be rewritten as
\begin{equation}
	\sum_{t_a\in\calM} t_a \frac{\pd}{\pd t_a} \F^{(h)}= (2-2h) \F^{(h)}
	\label{eq:HE2} 
\end{equation}

\subsection{Inversion of the loop insertion operator}

Let us compute here $\F^{(h)}$ for all $h>1$. $\F^{(0)}$ and $\F^{(1)}$ will be computed using other techniques. 
Note that from $\F^{(h)}$ and $\w_1^{(h)}$ are related by
\begin{equation}
	\w_1^{(h)}(p)=-\frac{\pd}{\pd V(p)} \F^{(h)}
	\label{eq:dVF} 
\end{equation} 
so in principle we only have to invert the loop insertion operator. We will propose an operator and act with it on $\w_1^{(h)}$, and then check that what we obtain satisfy \eqref{eq:dVF}.

\subsection{The \tops{$\H$}{H.} operator}

Following \cite{EyOr-05.1,EyOr-07.1} consider the integral operator acting on a differential $\phi(p)$ (recall the notation of (\ref{tableint})):
\begin{equation}
		\H \phi(\cdot):= -\sum_{t_a\in\calM} t_a \mathcal J_a(\phi) 
	\label{eq:Hop} 
\end{equation} 
As remarked in (\ref{ydxalt}) the choice $\phi(p)=B(p,q)$ yields $ -y\d x$.
If we perform an appropriate dissection $\mathcal D$ of $\Sigma$ along the basic cycles and along paths joining the points above infinity then we can write $y\d x = \d \Psi$ for a holomorphic $\Psi$ on $\mathcal D$. We can then rewrite (\ref{eq:Hop}) by integration by parts (basically a form of Riemann--bilinear identity)
\begin{equation}
	\begin{split}
		\H \phi(\cdot) & =
		-\Res_{q\to\text{Poles of $\phi(q)$}} \phi(q) \int_{r=o}^{r=q} y(r)\d x(r)
		= -\Res_{q\to\text{Poles of $\phi(q)$}} \phi(q) \Psi(q)\\
	\end{split}
\end{equation} 

Again in \cite{EyOr-07.1} the authors proved that
\begin{equation}
	(2-k-2h)\w_k^{(h)}(p_\K)=-\H \w_{k+1}^{(h)}(p_\K,\cdot)=\Res_{q\to\text{Poles of $\omega^{(h)}_{k+1}(q)$}} \w_{k+1}^{(h)}(q) \Psi(q)=\sum_i\Res_{q\to\alpha_i} \w_{k+1}^{(h)}(q) \Psi(q)
\label{eq:HOP2}
\end{equation} 
for all $(h,k)$ such that $2h+k-2>0$, with $\alpha_i$ being the branch points (only poles of $\w_{k+1}^{(h)}$).
Note that the $h=0,k=1$ case amounts to \eqref{ydxalt}.
The proof of this relation is obtained by applying the $\H$ operator into the recursive definition of $\w_k^{h}$. For more details see \cite{EyOr-07.1}.

\subsection{Computing \tops{$\F^{(h)}$}{Fh}}

Define the following infinite hierarchy of functions on the moduli space  $\calM$
\begin{equation}
	\tF^{(h)}=-\frac{\H\w_1^{(h)}(\cdot)}{2-2h}=\frac{\sum_i \Res_{q\to\alpha_i} \w_1^{(h)}(q)}{2-2h}
\end{equation} 
for $h>1$. Our claim is that $\tF^{(h)}=\F^{(h)}$. As a first check we apply $\d x(p)\frac{\pd}{\pd V(p)}$ to $\tF^{(h)}$,
\begin{equation}
	\begin{split}
		(2-2h)\d{x}(p)\frac{\pd}{\pd V(p)} \tF^{(h)} &= -\d{x}(p)\frac{\pd}{\pd V(p)} 
		\sum_i\Res_{q\to\alpha_i} \Psi(q)\w_1^{(h)}(q)\\
		&=-\sum_i\Res_{q\to\alpha_i} \left(\d{x}(p)\frac{\pd}{\pd V(p)} \Psi(q)\right)\w_1^{(h)}(q)
		-\sum_i\Res_{q\to\alpha_i} \Psi(q)\left(\d{x}(p)\frac{\pd}{\pd V(p)} \w_1^{(h)}(q)\right)\\
	\end{split}
\label{eq:dVtF} 
\end{equation} 
Let us work out the first term
\begin{equation}
	\begin{split}
		\d{x}(p)\frac{\pd}{\pd V(p)} \Psi(q)&=\int_o^q \d{x}(p)\frac{\pd}{\pd V(p)} y(r)\d x(r) \\
		&= - \int_{o}^{q} \left(B(p,r)-\frac{1}{3}\frac{\d x(p)\d x(r)}{(x(p)-x(r))^2}\right)\\
		&= - \d S_{q,o}(p) + \frac{\d{x}(p)}{3}\left(\frac{1}{x(p)-x(q)}-\frac{1}{x(p)-x(o)}\right)
	\end{split}
\end{equation} 
then we have
\begin{equation}
	\begin{split}
		&\sum_i\Res_{q\to\alpha_i} \left(\frac{\pd}{\pd V(p)} \Psi(q)\right)\w_1^{(h)}(q)\\
		&=- \w_1^{(h)}(p) - \sum_{i=0}^2 \w_1^{(h)}(p^{(i)})\\
		&=- \w_1^{(h)}(p).
	\end{split}
\end{equation} 
after moving contours and integrating.
The second term of \eqref{eq:dVtF} reads
\begin{equation}
	\begin{split}
		&\sum_i\Res_{q\to\alpha_i} \Psi(q)\d{x}(p)\frac{\pd}{\pd V(p)} \w_1^{(h)}(q)\\
		&=\sum_i\Res_{q\to\alpha_i} \Psi(q) \w_2^{(h)}(q,p)\\
		&=-(1-2h) \w_1^{(h)}(p)
	\end{split}
\end{equation} 
by eq. \eqref{eq:HOP2}.
Together we have
\begin{equation}
	\d{x}(p)\frac{\pd}{\pd V(x)} \tF^{(h)} = - \w_1^{(h)}
	\label{eq:dVF2} 
\end{equation} 
This is telling us that $\tF^{(h)}-\F^{(h)}$ may only depend on $T$ and the filling fractions.

In fact we can work out any variation with respect to the parameters $t_a\in\calM$, 
\begin{equation}
	\begin{split}
		(2-2h)\frac{\pd}{\pd t_a}\tF^{(h)}&= \sum_i \Res_{p\to\alpha_i} \left(\frac{\pd}{\pd t_a} \Psi(p)\right) \w_1^{(h)}(p)
		\sum_i \Res_{p\to\alpha_i} \Psi(p)\left(\frac{\pd}{\pd t_a} \w_1^{(h)}(p)\right) \\
		&=\sum_i \Res_{p\to\alpha_i} \left(\int_{r=o}^{r=p} \mathcal J_a(B(r,\cdot)) \w_1^{(h)}(p)\right)
		+\sum_i \Res_{p\to\alpha_i} \Psi(p) \mathcal J_a \left(\w_2^{(h)}(p,\cdot)\right) \\
		&=\mathcal J_a\left( \sum_i \Res_{p\to\alpha_i} \d S_{p,o}(\cdot)\w_1^{(h)}(p)\ri) 
		+\mathcal J_a\left( \sum_i \Res_{p\to\alpha_i} \Psi(p) \w_2^{(h)}(p,\cdot)\ri) \\
		&=\mathcal J_a\left(\w_1^{(h)}\ri)
		+ (1-2h)\mathcal J_a(\w_1^{(h)})\\
		&=(2-2h)\mathcal J_a\left(\w_1^{(h)}\ri)\\
		\Rightarrow\frac{\pd}{\pd t_a}\tF^{(h)}&=\mathcal J_a( \w_1^{(h)})
	\end{split}
\label{eq:dtFh}
\end{equation} 
We are allowed to change the order of the integrations because the integration path never intersect. We use \eqref{eq:HOP2} to work out the contribution from the second term.

It is clear now that $\tF^{(h)}$ satisfy homogeneity equation \eqref{eq:HE2}. Indeed, using the previous result we get
\begin{equation}
	\begin{split}
		\sum_{t_a\in\calM} t_a\frac{\pd}{\pd t_a} \tF^{(h)} = &
		\sum_{t_a\in\calM} t_a \mathcal J_a(\w_1^{(h)})\\
		=& - \H \w_1^{(h)}(\cdot) = (2-2h) \tF^{(h)}
	\end{split}
\end{equation} 
that together with \eqref{eq:dVF2}, shows that $\tF^{(h)}=\F^{(h)}$,
\begin{equation}
	\F^{(h)}=\frac{1}{2-2h}\sum_{i}\Res_{p\in\alpha_i} \Psi(p) \w_1^{(h)}(p)
	\label{eq:Fhf} 
\end{equation} 
with $\alpha_i$ the branch points and $\d \Psi(p)=y(p)\d x(p)$.

\subsection{Computing \tops{$\F^{(0)}$}{F0} and \tops{$\F^{(1)}$}{F1}}

This method determines all terms in the topological expansion of $\F$ except for $\F^{(0)}$ and $\F^{(1)}$. These two terms must be computed using different techniques. We do so in this section

\subsubsection{Formula for  \tops{$\F^{(0)}$}{F0}}
The computation of the planar limit of the free energy is  contained in \cite{Bert-06.1} and we report only the final formula (only minor notational differences occur).
We are seeking a function on the moduli space such that 
\begin{eqnarray}
\partial_{t_{k}^{(1)}} \F^{(0)} &=&  \left(\frac 2 3\Res_{\infty_1} - \frac 1 3 \Res_{\infty_0, \infty_2} \right) x^{k} y \d x\\
\partial_{t_{k}^{(2)}} \F^{(0)} &=& -\left(\frac 2 3\Res_{\infty_2} - \frac 1 3 \Res_{\infty_0, \infty_1} \right)(- x)^{k} y \d x\\
\partial_{\epsilon_j} \F^{(0)} &=& \oint_{\mathcal B_j} y\ d x\\
\partial_{t_{-1}^{(\nu)}} \F^{(0)}&=& \mu^{(\nu)} := (-)^\nu \slint_{\infty_0}^{\infty_\nu} y\d x\\
\partial_T \F^{(0)} &=& \slint_{\infty_2}^{\infty_1} y \d x
\end{eqnarray}
\begin{theorem}[Theorem 3.2 in \cite{Bert-06.1}]
\label{F0}
The planar limit of the free energy is given by the formula  (refer to (\ref{tableint}) for the meaning of the symbols)
\begin{eqnarray}
2 \mathcal F & =& \sum_{k=0}^{d_{1}} t_{k}^{(1)} 
\mathcal J_{t_k^{(1)}}(y\ d x) +   \sum_{j=0}^{d_{2}} t_{j}^{(2)}
\mathcal J_{t_{j}^{(2)}} (y\ d x) +T \,\mathcal J_T( y\ d x)  +\sum_{j=1}^g \sum_{k=1}^2 \epsilon_j^{(k)}  \mathcal J_{\epsilon_j^{(k)}} (y\ d x)=\\
&=& \sum_a t_a \mathcal J_a(y \d x)  = \sum_{a,b} t_a t_b \mathcal J_a\otimes \mathcal J_b(B)\label{freeenergy}
\end{eqnarray}
where
the notation $\mathcal J_a\otimes \mathcal J_b(\square)$ stands for the integral operator applied to a bi-differential on the two variables.
The last expression comes using (\ref{ydxalt}).
\end{theorem}

\subsubsection{Formula for \tops{$\F^{(1)}$}{F1}}

The first topological correction to the free energy can be computed adapting already standard methods (see for example \cite{EyKoKo-04.1,EyKoKo-05.1}) to our case. We present here the result and derive it in appendix \ref{App:F01}.

\begin{equation}
	\begin{split}
		\F^{(1)}=-\frac{1}{24}\ln\left({\tau_{B x}}^{12}\, \prod_{\alpha_i}y^{\prime}(\alpha_i)\right)
	\end{split}
\end{equation}

\appendix

\section{Loop equations and spectral curve}\label{App:CoV}

In this appendix we derive the master loop equations, that will allow us to write a recursive formula for the topological expansion of the Cauchy matrix model.

One way to find loop equations is by applying infinitesimal changes of variables to the matrix integral, and looking at the first order variation of the integral, which must vanish.

\subsection{Loop equation: General form}\label{sSec:LE:GF}
The loop equations are a glorified version of the following fact:
suppose we have a multivariate integral on a domain $\Omega$,   $\int_\Omega f(\vec x) \d  x$ and a vector field $\dot {\vec x} = \vec h(\vec x)$. 
Then the action of the vector field on the integrand induces the integral of a ``total derivative'', namely
\be
\frac {\d {}}{\d t}   \int_{\Omega} f (\vec x)\d{} ^n x = \int_{\Omega}\left( \vec h \cdot \nabla f +f  {\rm div} (\vec h)\ri)\d{}^n x = \int_{\Omega} {\rm div}\left(f \vec h \ri) = -\int_{\pa \Omega} f(\vec x) \vec h(\vec x) \d{}^{n-1} x
\ee 
where the last integral is the ``flux'' of the vector on the boundary of $\Omega$. If either $f$ or $\vec h$ vanish on (or are tangential to) the boundary, then the variation is zero and thus yields identities amongst different integrals. In our case $\vec  x$ are the two matrices and the domain of integration $\Omega$ is the cone of positive matrices. 
Consider the vector field induced by the infinitesimal change of variable
\be
M_j \to M_j+\ep \delta M_j\ .\label{vectf}
\ee
The partition function $\ZC$ is an integral on the space of positive definite matrices; the infinitesimal variation of the integrand under (\ref{vectf}) is the divergence of a vector
\begin{equation}
\begin{split}
M_1 &\to M_1+\ep \delta M_1\\
\ZC &\to \ZC+\ep \delta \ZC + \OO{\ep^2},
\end{split}
	\label{eq:LE1}
\end{equation}
In order to have a vanishing variation the vector field must vanish on the set of matrices with at least one zero-eigenvalue. Under this condition we will have  $\delta \ZC=0$.
From the explicit form of $\ZC$ \eqref{eq:CMM}, $\delta \ZC=0$ is expressed as
\begin{equation}
	\BK{J(\delta M_1)}-\frac{N}{T}\BK{\tr{\delta M_1 V_1^\prime(M_1)}}
	-\BK{\Tr{1}{\delta M_1 \Tr{2}{\frac{1}{M_1\otimes\id +\id\otimes M_2}}}}=0.
	\label{eq:LE2}
\end{equation}
$J(\delta M_1)$ is the factor coming from the Jacobian of the change of variables. This is the trace of the matrix $\frac{\pd{} (\delta M_1)_{i,j}}{\pd{} (M_1)_{k,l}}$, i.e.
\begin{equation}
	J(\delta M_1)=\sum_{i,j=1}^N\frac{\pd{} (\delta M_1)_{i,j}}{\pd{} (M_1)_{i,j}}
	\label{eq:Jac}
\end{equation}
For our purposes\footnote{we leave it to the reader to check these rules. Also, see \cite{Ey-03.1}.} this computation reduces to two simple rules
\begin{equation}
\begin{split}
J\left(\frac{1}{\xi-M_1}A\right)&=\tr{\frac{1}{\xi-M_1}A}\tr{\frac{1}{\xi-M_1}}\\
J\left(\tr{\frac{1}{\xi-M_1}A}B\right)&=\tr{\frac{1}{\xi-M_1}A\frac{1}{\xi-M_1}B}
\end{split}
\label{eq:SandM}
\end{equation}
where $A$ and $B$ are in general combinations of constant matrices in $M_1$ and factors of the type $\frac{1}{\xi-M_1}$. Of course, if $A$ and $B$ do contain such {\it resolvent-like} terms, the Leibniz rule applies to $A$ and $B$.

\subsection{Master Loop Equations}\label{sSec:MLE}

In the Cauchy model two loop equations appear to be crucial. One of them is quadratic in the resolvents and the other is cubic. In this section we derive both of them before extracting all information in the following sections.

\subsubsection{{Quadratic Master Loop Equation}}\label{ssSec:QMLE}

\noindent Consider the following change of variables for $M_1$
\begin{equation}
	\delta M_1=\frac{T^2}{N^2}\left(\frac{1}{\xi-M_1}-\frac{1}{\xi}\right).
	\label{eq:MLE.1}
\end{equation}
Note that the the vector field above is permissible since it preserves the cone of positive matrices. Indeed any zero eigenvalue of $M_1$ is  invariant under the flow; the eigenvalues of $M^\prime=M+\ep \frac{T^2}{N^2}\left(\frac{1}{\xi-M}-\frac{1}{\xi}\right)$ are 
$m_i^\prime=m_i+\ep\frac{T^2}{N^2}\left(\frac{1}{\xi-m_i}-\frac{1}{\xi}\right)\stackrel{m_i\to 0}{\longrightarrow}0$.
Using the rules (\ref{eq:SandM})  we find
\begin{equation}
	J(\delta M_1)=\left(\frac{T}{N}\tr{\frac{1}{\xi-M_1}}\right)^2.
	\label{eq:MLE.2}
\end{equation}
The associated loop equation is
\begin{equation}
	\BK{\left(W_1(\xi)\right)^2}-V_1^\prime(\xi)\BK{W_1(\xi)}+\BK{P_1(\xi)}+\BK{\Xi_1(\xi)}
	+\frac{1}{\xi}\BK{I_1}=0.
	\label{eq:MLE.3}
\end{equation}
where we have defined
\begin{equation}
\begin{split}
	W_1(\xi)&=\frac{T}{N}\tr{\frac{1}{\xi-M_1}}\\
	P_1(\xi)&=\frac{T}{N}\tr{\frac{V_1^\prime(\xi)-V_1^\prime(M_1)}{\xi-M_1}}\\
	\Xi_1(\xi)&=-\frac{T^2}{N^2}\Tr{1}{\frac{1}{\xi-M_1}\Tr{2}{\frac{1}{M_1\otimes\id +\id\otimes M_2}}}\\
	I_1&=\frac{T}{N}\tr{V_1^\prime(M_1)}+\frac{T^2}{N^2}\Tr{1,2}{\frac{1}{M_1\otimes\id +\id\otimes M_2}}.
\end{split}	
	\label{eq:Def.1}
\end{equation}
Make a similar variation for $M_2$
\begin{equation}
	\delta M_2=\frac{T^2}{N^2}\left(\frac{1}{\eta-M_2}-\frac{1}{\eta}\right)
	\label{eq:MLE.4}
\end{equation}
with the Jacobian
\begin{equation}
	J(\delta M_2)=\left(\frac{T}{N}\tr{\frac{1}{\eta-M_2}}\right)^2
	\label{eq:MLE.5}
\end{equation}
with the associated loop equation
\begin{equation}
	\BK{\left({W}_2(-\eta)\right)^2}-V_2^\prime(\eta)\BK{{W}_2(-\eta)}
	+\BK{{P}_2(-\eta)}+\BK{{\Xi}_2(-\eta)}
	+\frac{1}{-\eta}\BK{{I}_2}=0
	\label{eq:MLE.6}
\end{equation}
and the definitions
\begin{equation}
\begin{split}
	{W}_2(-\eta)&=\frac{T}{N}\tr{\frac{1}{\eta-M_2}}\\
	{P}_2(-\eta)&=\frac{T}{N}\tr{\frac{V_2^\prime(\eta)-V_2^\prime(M_2)}{\eta-M_2}}\\
	{\Xi}_2(-\eta)&=-{\frac{T^2}{N^2}}\Tr{2}{\frac{1}{\eta-M_2}\Tr{1}{\frac{1}{M_1\otimes\id +\id\otimes M_2}}}\\
	{I}_2&=-\left(\frac{T}{N}\tr{V_2^\prime(M_2)}
	+{\frac{T^2}{N^2}}\Tr{1,2}{\frac{1}{M_1\otimes\id +\id\otimes M_2}}\right).
\end{split}
	\label{eq:Def.2}
\end{equation}
Combining both variations, setting $\eta=-\xi=x$ and using the relation
\begin{equation}
	{\Xi}_1(x)+\Xi_2(x)=W_1(x){W}_2(x)
	\label{eq:rel}
\end{equation}
we obtain the {\it quadratic master loop equation}
\begin{equation}
\begin{split}
	&\BK{\left(W_1(x)\right)^2}+\BK{\left({W}_2(x)\right)^2}+\BK{W_1(\x){W}_2(\x)}
	-V_1^\prime(\x)\BK{W_1(\x)}-V_2^\prime(-\x)\BK{{W}_2(\x)}\\
	&\hspace{80pt}+\BK{P_1(\x)}+\BK{{P}_2(\x)}+\frac{1}{\x}\left(\BK{I_1}+\BK{{I}_2}\right)
	=0.
\end{split}
	\label{eq:QMLE}
\end{equation}
Using the functions $Y_i$ defined in \ref{sSec:Def}
the equation \eqref{eq:QMLE} is expressed in a much more compact way
\begin{equation}
	\BK{\left(Y_1(\x)\right)^2}+\BK{\left({Y}_2(\x)\right)^2}+\BK{Y_1(\x){Y}_2(\x)}
	=\BK{R(\x)}-\frac{1}{\x}\BK{I_1+{I}_2}:=\BK{\hat{R}(\x)}
	\label{Aeq:YQMLE}
\end{equation}
where
\begin{equation}
\begin{split}
	R(\x)&=\frac{1}{3}(V_1^\prime(\x)^2+V_2^\prime(-\x)^2-V_1^\prime(\x)V_2^\prime(-\x))-P_1(\x)-{P}_2(\x)\\
	      &=(U^\prime_1(\x))^2+(U^\prime_2(\x))^2+U^\prime_1(\x)U^\prime_2(\x)-P_1(\x)-{P}_2(\x)
\end{split}
	\label{Aeq:RDef}
\end{equation}
is, in the limit $N\to\infty$, the same $R(x)$ found in \cite{BeGeSz-08.2}.

\Remark{Note that in \cite{BeGeSz-08.2} the $\frac{1}{\x}$ term in the RHS of \eqref{eq:RDef} is absent due to the constraints used on the potential which guaranteed the support of the equilibrium measures to be disjoint from $0$. In this case one can verify that the expectation of $I_1+I_2$ is a total derivative vanishing on the boundary of the integration and hence it is zero.}

\Remark{Note that $R(\x)$ is a rational function in the variable $\x$ with only one pole at $x=0$ of order $2$ coming from the squares of $V_k^\prime$. Clearly, although each term on the LHS has cuts along the real axis $\R$, their combination is rational function with only one pole at $\xi=0$ of order $2$ in general, or of order $1$ in case $t_{-1}^{(k)}=0$.}

\subsubsection{Cubic Master Loop Equation}\label{ssSec:CMLE}

For the {\it cubic} master loop equation the procedure is exactly the same but with slightly more complicated changes of variables. We are going to perform one change of variables for each $M_1$ and $M_2$, and then combine them.

Consider the following change of variables $\delta M_1$ and its related Jacobian $J(\delta M_1)$
\begin{equation}
\begin{split}
	\delta M_1&=\frac{T^2}{N^2}\left(\frac{1}{\xi-M_1}-\frac{1}{\xi}\right)
	\left(\frac{T}{N}\tr{\frac{1}{\xi-M_1}}-\frac{T}{N}\tr{\frac{1}{\eta-M_2}}-V_1^\prime(\xi)+V_2^\prime(\eta)\right)\\
	J(\delta M_1)&=\left(\frac{T}{N}\tr{\frac{1}{\xi-M_1}}\right)^3
	-\left(\frac{T}{N}\tr{\frac{1}{\xi-M_1}}\right)^2\frac{T}{N}\tr{\frac{1}{\eta-M_2}}
	-(V_1^\prime(\xi)-V_2^\prime(\eta))\left(\frac{T}{N}\tr{\frac{1}{\xi-M_1}}\right)^2\\
	&\qquad +{\frac{T^3}{N^3}}\tr{\frac{1}{\left(\xi-M_1\right)^3}} 
	-{\frac{T^3}{N^3}}\frac{1}{\xi}\tr{\frac{1}{\left(\xi-M_1\right)^2}}
\end{split}
	\label{eq:CMLE.1}
\end{equation}
after a little bit of algebra one gets the following loop equation
\begin{equation}
\begin{split}
	0=&\BK{W_1(\xi)^3}
	+\frac{T^2}{N^2}\left(\frac{1}{2}\frac{\d{}^2}{\d{\xi}^2}+\frac{1}{\xi}\frac{\d{}}{\d{\xi}}\right)\BK{W_1(\xi)}-
	\BK{W_1^2(\xi){W}_2(-\eta)}\\
	&\hspace{40pt}-(2V_1^\prime(\xi)-V_2^\prime(\eta))\BK{W_1(\xi)^2}
	+V_1^\prime(\xi)\BK{W_1(\xi){W}_2(-\eta)}\\
	&\hspace{80pt}+\BK{\Xi_1(\xi)\left(W_1(\xi)-{W}_2(-\eta)-V_1^\prime(\xi)+V_2^\prime(\eta)\right)}\\
	&\hspace{120pt} +\BK{\left(P_1(\xi)+\frac{1}{\xi}I_1\right)
	\left(W_1(\xi)-{W}_2(-\eta)-V_1^\prime(\xi)+V_2^\prime(\eta)\right)}
\end{split}
	\label{eq:CMLE.2}
\end{equation}

Consider also the symmetric change of variable for $M_2$
\begin{equation}
\begin{split}
	\delta M_2&= {\frac{T^2}{N^2}}\left(\frac{1}{\eta-M_2}-\frac{1}{\eta}\right)
	\left(\frac{T}{N}\tr{\frac{1}{\eta-M_2}}-\frac{T}{N}\tr{\frac{1}{\xi-M_1}}-V_2^\prime(\eta)+V_1^\prime(\xi)\right)\\
	J(\delta M_2)&=\left(\frac{T}{N}\tr{\frac{1}{\eta-M_2}}\right)^3
	-\left(\frac{T}{N}\tr{\frac{1}{\eta-M_2}}\right)^2\frac{T}{N}\tr{\frac{1}{\xi-M_1}}
	-(V_2^\prime(\eta)-V_1^\prime(\xi))\left(\frac{T}{N}\tr{\frac{1}{\eta-M_2}}\right)^2\\
	&\qquad +{\frac{T^3}{N^3}}\tr{\frac{1}{\left(\eta-M_2\right)^3}} 
	-\frac{T^3}{N^3}\frac{1}{\eta}\tr{\frac{1}{\left(\eta-M_2\right)^2}}
\end{split}
	\label{eq:CMLE.3}
\end{equation}
to obtain
\begin{equation}
\begin{split}
	0=&\BK{{W}_2(-\eta)^3}
	+{\frac{T^2}{N^2}}\left(\frac{1}{2}\frac{\d{}^2}{\d{(-\eta)}^2}+\frac{1}{-\eta}\frac{\d{}}{\d{(-\eta)}}\right)
	\BK{{W}_2(-\eta)}-
	\BK{{W}_2^2(-\eta)W_1(\xi)}\\
	&\hspace{40pt}+(V_1^\prime(\xi)-2V_2^\prime(\eta))\BK{{W}_2(-\eta)^2}
	+V_2^\prime(\eta)\BK{{W}_2(-\eta)W_1(\xi)}\\
	&\hspace{80pt}-\BK{{\Xi}_2(\xi)\left({W}_2(-\eta)
	-W_1(\xi)-V_2^\prime(\eta)+V_1^\prime(\xi)\right)}\\
	&\hspace{120pt} +\BK{\left(P_2(-\eta)+\frac{1}{-\eta}I_2\right)
	\left({W}_2(-\eta)-W_1(\xi)-V_2^\prime(\eta)+V_1^\prime(\xi)\right)}
\end{split}
	\label{eq:CMLE.4}
\end{equation}

Now, set $-\eta=\xi=x$, subtract equation \eqref{eq:CMLE.4} from \eqref{eq:CMLE.2}, use \eqref{eq:rel} and introduce $Y_1$ and ${Y}_2$ to obtain the {\it cubic master loop equation}
\begin{equation}
\begin{split}
	\BK{Y_1(\x)^3}-\BK{{Y}_2(\x)^3}-\BK{\hat{R}(\x)(Y_1(\x)-{Y}_2(\x))}
	-\frac{T^2}{N^2}\left(\frac{1}{2}\frac{\d{}^2}{\d{\x}^2}+\frac{1}{\x}\frac{\d{}}{\d{\x}}\right)
	\BK{W_1(\x)-{W}_2(\x)}=0
\end{split}
	\label{eq:YCMLE}
\end{equation}
This equation implies that both $Y_1(\x)$ and ${Y}_2(\x)$ satisfy the same equation
\begin{equation}
\begin{split}
	\BK{Y_k(\x)^3}-\BK{\hat{R}(\x)Y_k(\x)}
	-{\frac{T^2}{N^2}}\left(\frac{1}{2}\frac{\d{}^2}{\d{\x}^2}+\frac{1}{\x}\frac{\d{}}{\d{\x}}\right)
	\BK{W_k(\x)}&=\BK{D(\x)}\,,\quad \text{for $k=1,2$}\\
\end{split}
	\label{Aeq:YCMLE.a}
\end{equation}
where $D(\x)$ is still unknown. To find an expression for $D(\x)$ we use the same basic variations above to obtain the following equation
\begin{equation}
	\BK{Y_0(\x)^3}-\BK{\hat{R}(\x)Y_0(\x)}
	-{\frac{T^2}{N^2}}\left(\frac{1}{2}\frac{\d{}^2}{\d{\x}^2}+\frac{1}{\x}\frac{\d{}}{\d{\x}}\right)
	\BK{W_0(\xi)}=-\BK{Y_1(\x)^2{Y}_2(\x)}-\BK{Y_1(\x){Y}_2(\x)^2}
	\label{eq:YCMLE.0}
\end{equation}
where we define
\begin{equation}
\begin{split}
	W_0(\x)&=-W_1(\x)-{W}_2(\x)\\
	Y_0(\x)&=-Y_1(\x)-{Y}_2(\x)=U^\prime_0(\x)-W_0(\x)
\end{split}
	\label{eq:YCMLE.0Def}
\end{equation}
On the other side, from the two cubic master loop equations \eqref{Aeq:YCMLE.a} we have
\begin{equation}
	\BK{Y_0(\x)^3}-\BK{\hat{R}(\x)Y_0(\x)}
	-{\frac{T^2}{N^2}}\left(\frac{1}{2}\frac{\d{}^2}{\d{\x}^2}+\frac{1}{\x}\frac{\d{}}{\d{\x}}\right)
	\BK{W_0(\x)}=-2\BK{D(\x)}-3\BK{Y_1(\x)^2{Y}_2(\x)}-3\BK{Y_1(\x){Y}_2(\x)^2}
	\label{Aeq:YCMLE.1}
\end{equation}
The RHS of equations \eqref{eq:YCMLE.0} and \eqref{Aeq:YCMLE.1} determine, after some more algebra and more loop equations,
\begin{equation}
\begin{split}
	\BK{D(\x)}&=-\BK{Y_1(\x)^2{Y}_2(\x)+Y_1(\x){Y}_2(\x)^2}\\
	&=-U^\prime_0(\x)U^\prime_1(\x)U^\prime_2(\x)-U^\prime_1(\x)\BK{{P}_2(\x)}-U^\prime_2(\x)\BK{P_1(\x)}
	-\frac{1}{\x}\left(U^\prime_1(\x)\BK{{I}_2}+U^\prime_2\BK{I_1}\right)\\
	&\qquad +\BK{Q_1(\x)+{Q}_2(\x)}+\BK{W_1(\x)^2{W}_2(\x)}+\BK{W_1(\x){W}_2(\x)^2}+\BK{S_1(\x)}+\BK{{S}_2(\x)}
\end{split}
	\label{Aeq:DDef}
\end{equation}
with
\begin{equation}
\begin{split}
	Q_1(\x)=&{\frac{T^2}{N^2}}\tr{\frac{V_1^\prime(\x)-V_1^\prime(M_1)}{\x-M_1}
	\Tr{2}{\frac{1}{M_1\otimes\id+\id\otimes M_2}}}\\
	{Q}_2(-\x)=&{\frac{T^2}{N^2}}\tr{\frac{V_2^\prime(\x)-V_2^\prime(M_1)}{\x-M_1}
	\Tr{1}{\frac{1}{M_1\otimes\id+\id\otimes M_2}}}\\
	S_1(\x)=&{\frac{T^2}{N^2}}\tr{\frac{V_1^\prime(M_1)}{\x-M_1}
	\Tr{2}{\frac{1}{M_1\otimes\id+\id\otimes M_2}}}\\
	{S}_2(-\x)=&{\frac{T^2}{N^2}}\tr{\frac{V_2^\prime(M_1)}{\x-M_1}
	\Tr{1}{\frac{1}{M_1\otimes\id+\id\otimes M_2}}}\\
\end{split}	
	\label{eq:DDefDef}
\end{equation}
In \cite{BeGeSz-08.2} the authors already considered the large $N$ limit of this equation and found the large $N$ limit of $\BK{D(x)}$ that we call $D^{(0)}(x)$. It is clear that the last $4$ terms in \eqref{Aeq:DDef} only contribute to the pole structure at $x=0$. The analysis in \cite{BeGeSz-08.2} shows that the pole coming from these terms is at most of order 2.
The actual order of the pole at $x=0$ depends on $t_{-1}^{(k)}$ through the products of $U^{\prime}_i(x)$.
Whenever both $t_{-1}^{(i)}>0$ are different, the pole is of order $3$ and comes from the product $U_0^\prime(x)U_1^\prime(x)U_2^\prime(x)$. Whenever the symmetry $t_{-1}^{(1)}=t_{-1}^{(2)}$ is present the pole is of order $2$ since $U_0$ does not have a log term. 

\Remark{Note that now, eq. \eqref{Aeq:YCMLE.a} is satisfied by $k=0,1,2$.}

\Remark{Note that, by construction, the LHS of equation \eqref{Aeq:YCMLE.a} may have cuts only on $\R_+$ for $k=1$, while for $k=2$ the cuts may only appear on $\R_-$. Indeed, $Y_1(x)$ has poles only the spectrum of $M_1$, $x=x_i\in\R_+$ ,while ${Y}_2(x)$ has poles only on the spectrum of $-M_2$, $x=-y_i\in\R_-$. Thus $\BK{D(x)}$ is a meromorphic function of $x\in\C$ everywhere except at $x=0$ where it may have poles.}

\section{Definitions on the algebraic curve}\label{App:AC}

In this appendix we will give a brief review of several definitions of objects on the algebraic curve.  The notions are relatively standard and we refer for example to \cite{FarkasKra, Faybook}.
\subsection{The algebraic curve}\label{sSec:AC}

Given an algebraic curve $E(x,y)=0$ (irreducible polynomial equation of degree $d_1$ in $x$ and $d_2$ in $y$), its points $(x,y)$ parametrize the points $p$ of a Riemann surface $\Sigma$. Choosing $x\in\C\cup\{0\}$ ($y\in\C\cup\{0\}$) as the base coordinate there are $d_2$ ($d_1$) solutions $y_i(x)$ ($x_i(y)$) of the algebraic curve. Those functions define $d_2$ $x$-sheets ($d_1$ $y$-sheets) which form a chart on the Riemann surface. 
Since, for every point $p\in\Sigma$ there is a pair $(x,y)$ that solve $E(x,y)=0$ we can define two meromorphic functions $x(p),y(p)$ on $\Sigma$.
For a generic $p=p^{(0)}\in\Sigma$ there are $d_2-1$ other points $p^{(i)},\,i=1,\dotsc,d_2-1$ such that $x(p^{(i)})=x(p)$. Each one of these points lies in a different $x$-sheet. In general, on these points we have $y(p^{(i)})\not=y(p^{(j)})$ for all pairs $i,j$.
The same definitions can be made based on $y$-sheets but we do not use it in this work.

\subsection{The branch points, branch cuts and locally conjugated points}\label{sSec:BP}

We will consider curves where $\d x(p)$ and $\d y(p)$ never vanish at the same point $p$ (i.e. without cusps).
The points where $\d x(p)=0$ are called branch points and are denoted by $\alpha_i$. 
We only  consider algebraic curves where the zeros of $\d x(p)$ are simple.
The sheets are connected pairwise by {\it cuts} that go from one branch point to another, every branch point having only one attached cut.

In a neighborhood $U_\alpha$  of any branchpoint $\alpha$ there are {\bf two} points $p, \overline{ p}\in \Sigma$ such that $x(p) = x(\overline {p})$. The map $\overline{ \phantom {-}}: U_\alpha \mapsto U_\alpha$ is  a local involution, namely,  $\overline{\overline{p}}=p$.
Note that the notion is purely local and has no intrinsic meaning at the global level. If one uses the local coordinate $z(p):= \sqrt{x(p)-x(\alpha)}$ as a conformal parameter for $U_\alpha$ then $z(\overline{p}) = -z(p)$.

\subsection{Cycles and genus}\label{sSec:CG}

In every Riemann surface we can choose a basis of cycles for the homology group. These cycles can be chosen to form a canonical basis, i.e. two types of cycles $\A_i,\B_i$ homologically independent which satisfy the intersection conditions
\begin{equation}
	\begin{split}
		\A_i\cdot\A_j&=0=\B_i\cdot\B_j\\
		\A_i\cdot\B_j&=-\B_j\cdot\A_i=\delta_{i,j}.
	\end{split}
	\label{eq:CCC}
\end{equation}  
A Riemann surface of genus $g$ has exactly $g$ pairs of conjugated cycles $\A_i,\B_i$.
Usually one defines the homotopy group by choosing related cycles $\overline{\A}$ and $\overline{\B}$. These cycles are equivalent to the previous ones with the constraint that all pass through a common point $p_0$. Cutting $\Sigma$ along $\overline{\A}$ and $\overline{\B}$ defines $\overline{\Sigma}$ which is isomorphic to an open region of the complex plane $\C$. This is called the canonical dissection of $\Sigma$

\subsection{Abelian differentials.}\label{sSec:AB}

There are three type of meromorphic differentials on an algebraic curve.
The {\bf differentials of the first kind} (holomorphic)  form a vector space of dimension $g$ and we denote them by $\d u_i$. We can always choose a basis $\d u_i(p)$ such that it satisfy
\begin{equation}
	\oint_{\A_j}\d u_i(p)=\delta_{i,j}
	\label{eq:AD1}
\end{equation}
With this normalization we define the Riemann period matrix as the matrix of $\B$ periods
\begin{equation}
	\tau_{i,j}=\oint_{\B_i}\d u_j(p)
	\label{eq:RPM}
\end{equation}
This matrix is symmetric and its imaginary part is positive definite $\Im(\tau)>0$.
These differentials $\d u_i(p)$ are called normalized Abelian differentials of the first kind.

The Abelian {\bf differentials of the second kind} are defined as meromorphic differentials with poles but with zero residues. A basis for these differentials may be called $\d\Omega_k^{(q)}(p)$ which is a meromorphic differential with a pole at $p=q$ without residue normalized as follows
\begin{equation}
	\begin{split}
		\d\Omega_k^{(q)}(p)&\sim (-kz_q(p)^{-k-1}+\OO{1})\d z_q(p)\\
		\oint_{\A_i} \d\Omega_k^{(q)}(p) & =0
	\end{split}
\end{equation}
in some local parameter $z_q(p)$ such that $z_q(q)=0$.

Finally the {\bf differentials of the  third kind} are meromorphic differentials that have only poles of first order. Again we will choose a basis for these differentials consisting of differentials $\d S_{q,r}(p)$ with only two poles of order one, say at $p=q,r$, with residue $1,-1$ respectively, and with vanishing $\A$ cycles, i.e. normalized following
\begin{equation}
	\begin{split}
		\d S_{q,r}(p)&\sim \left(\frac{1}{z_q(p)}+\OO{1}\right)\d z_q(p)\\
		\d S_{q,r}(p)&\sim \left(\frac{-1}{z_r(p)}+\OO{1}\right)\d z_r(p)\\
		\oint_{\A_i} \d S_{q,r}(p) & =0
	\end{split}
\end{equation}
where the local coordinates are defined as above.
The differentials in this basis is called normalized Abelian differentials of the third kind.
\subsection{The fundamental bi-differential and relations with the Abelian differentials}

We also introduce the fundamental bi-differential (Bergmann kernel) $B(p,q)$ which is a meromorphic bi-differential with poles only at $p=q$ of order two with zero residue, and normalized in the following way
\begin{equation}
	\begin{split}
		B(p,q)&\sim \frac{\d z(p)\d z(q)}{(z(p)-z(q))^2}+\OO{1}\\
		\oint_{\A_i}B(p,q)&=0 \qquad \text{for $i=1,\dotsc,g$}
	\end{split}
\end{equation}
this bi-differential is fundamentally connected with the Abelian differentials presented above.
\begin{equation}
	\begin{split}
		B(p,q)&=B(q,p)\\
		\d f(p)&=\Res_{q\to p}B(p,q) f(q)\\
		\sum_{i}B(p^{(i)},q)&=\frac{\d x(q)\d x(p)}{(x(p)-x(q))^2}\\
		\oint_{\B_j}B(p,\cdot)&=2\pi i \d u_j(p)\\
		\d S_{q,r}(p)&=\int_r^q B(p,\cdot)\\
		\d \Omega_k^{(r)}(p) &= \Res_{q\to p} B(p,q) z_r(q)^{-k}=-\Res_{q\to r} B(p,q) z_r(q)^{-k}\\
		\oint_{\B_j}\d S_{q,r}(\cdot)&=2\pi i \int_r^q \d u_j(\cdot)\\
		\oint_{t=p}^{t=q} \d S_{s,r}(t) &= \oint_{t=r}^{t=s} \d S_{q,p}(t) 
	\end{split}
	\label{eq:DifProp} 
\end{equation}

In all these equations, the integrations paths that are not closed cycles are constrained to not cross the $\A$ and $\B$ cycles, in other words belongs to the cut algebraic curve along the cycles.

\section{Derivation of \tops{$\F^{(1)}$}{F1}}\label{App:F01}

For the reader's convenience we derive the first topological correction $\F^{(1)}$ following  \cite{EyKoKo-04.1,EyKoKo-05.1}.

The computation consists in integrating the relation 
\begin{equation}
	\d x(p)\frac{\pd}{\pd V(p)} \F^{(1)} = -\w_1^{(1)} \label{strong}
\end{equation} 
The equation \eqref{strong} should be taken as a generating function for the derivatives of $\F^{(1)}$ with respect to the moduli in the potentials with $x(p)^{-1}$ as the formal expansion parameter. The equivalent formula for general moduli (say $T$, $\eta$'s and $\epsilon$'s) is
\be
\pa_{t_a} \F^{(1)} = -\mathcal J_a (\w_1^{(1)})
\ee 
with $\mathcal J_a$ read off table (\ref{tableint}).

The differential   $\w_1^{(1)}$ is obtained from equation \ref{eq:Sol.a}
\begin{equation}
	\begin{split}
		\w_1^{(1)} &= -\sum_{\alpha} \Res_{p\to\alpha} \frac{1}{2}\frac{\d S_{p,\bar{p}}(q)}{(y(p)-y(\bar{p}))\d x(p)}
		B(p,\bar{p})
	\end{split}
\end{equation} 
The sum extends over all branchpoints, namely, the zeroes of $\d x(p)$. 
The residue around the branchpoint $\alpha$  is computed straightforwardly by Taylor-expanding the expression.
After a computation in the local parameter $z_\alpha := \sqrt{x(p)-x(\alpha)}$ one obtains
\begin{equation}
	\begin{split}
		\w_1^{(1)}(q)=\sum_{\alpha} \left(\frac{B^{\prime\prime}(q,\alpha)}{96 y^\prime(\alpha)}
		+B(q,\alpha)\left(\frac{S_B(\alpha)}{24 y^\prime(\alpha)}
		-\frac{y^{\prime\prime\prime}(\alpha)}{96(y^\prime(\alpha))^2}\right)\right)\\
		=\underbrace{\sum_{\alpha} B(q,\alpha)\frac{S_B(\alpha)}{24 y^\prime(\alpha)}}_{(A)} + 
		\underbrace{\sum_{\alpha} \left(\frac{B^{\prime\prime}(q,\alpha)}{96 y^\prime(\alpha)}
		-B(q,\alpha)\frac{y^{\prime\prime\prime}(\alpha)}{96(y^\prime(\alpha))^2}\right)}_{(B)}
	\end{split}
\label{eq:w11}
\end{equation} 
We have emphasized the two expression above because each of them is the variation of a separate term, as we now explain.
We first need the 
\begin{lemma}

\label{lemmavars}
Let $\pa_{t}$ denote any variation of the moduli of the curve and recall the correspondence (\ref{tableint}). Let $a = x(\alpha)$, $\d x(\alpha)=0$ be any of the critical values  and denote the jet-expansion of $y(z)$ as $y(z) = y(\alpha)+ y'(\alpha) z + \dots$. Then 
\bea
\pa_{t} a& =& -\frac 1{y'(\alpha)}\mathcal J_t \left(B(\cdot,\alpha)\right)\\
\pa_{t} y^{(k)}(\alpha) &=& \frac 1{2(k+1)} \mathcal J_t \left(B^{(k+1)}(\cdot,\alpha) - \frac{y^{(k+2)}(\alpha)}{y'(\alpha)} B(\cdot,\alpha)  \ri)
\eea
where the evaluation of the bidifferential and the derivatives are done in the local parameter $z= \sqrt{x-a}$.
\end{lemma}
{\bf Proof.} By the discussion preceding (\ref{tableint}) we have
\be
\pa_{t}  y(p) \d x(p) = \mathcal J_t \left(B(\cdot ,p) \ri)
\ee
Consider the jet expansion of the above identity near $p=\alpha$ in the coordinate $z= z_\alpha = \sqrt {x-a}$. Recall that the variations are taken at fixed $x$--value but the critical value $a$ is not fixed and hence $\pa_t z = -\frac {\pa_t a}{2z}$. With this in mind we have
\begin{equation}
\begin{split}
\pa_{t} y(p) \d x(p)  &= \pa_{t} (y(z))  \d x =  \left( (\pa_{t} y)_{|_z}(z) + y'(z) \pa_{t} z \ri) \d x = \\
&= \left((\pa_{t} y)(\alpha) +( \pa_{t}  y')(\alpha) z + (\pa_{t} y'')(\alpha) \frac {z^2}2 + \dots 
-\frac {\pa_{t} a}{2 z} \left(y'(\alpha)   + y''(\alpha) z + \frac {y'''(\alpha)} 2 z^2 + \dots \ri) \ri) 2 z \d z\\
& =\mathcal J_t  \left( B(\cdot ,\alpha) + B'(\cdot ,\alpha) z + B''(\cdot,\alpha) \frac {z^2}2 + \dots\ri)
\end{split}
\end{equation}
where the primes denote derivatives w.r.t. $z$ (applied to the second variable in the case of the bidifferential above).
Comparing the coefficients of the jet-expansion in $z$ we find 
\bea
\pa_t a &=& - \mathcal J_t \left(\frac {B(\cdot ,\alpha)}{y'(\alpha)}\right) \\
\frac {2}{k!}\pa_t y^{(k)}(\alpha) - \frac 1{(k+1)!}\,y^{(k+2)}(\alpha)  \pa_t a &=& \frac 1{(k+1)!} \mathcal J_t\left( B^{(k+1)}(\cdot ,\alpha) \right )
\eea
which immediately yields the assertion. {\bf Q.E.D.}\par \vskip 5pt 
Using Lemma (\ref{lemmavars}) with $k=1$ on the term (B) in (\ref{eq:w11}) we see that 
\be
(B) = \frac 1{24} \d{x}(p)\frac{\pa}{\pa{V(q)}} \ln \prod_j y'(\alpha_j)
\ee
To identify (A) we need to resort to the results of \cite{KokotovKorotkin1} where (paraphrasing) it was shown that 
\begin{proposition}(\cite{KokotovKorotkin1})
Suppose $x:\Sigma \to \mathbb P^1$ is a branched covering of the Riemann--sphere with branchpoints $\alpha_j$. Then the following differential on the space of critical values of such coverings is closed
\be
\pa_t \ln\tau_{Bx}:=\sum_j \frac{\pa}{\pa a_j}(\ln  \tau_{Bx}) \pa_t a_j := -\frac{1}{12} \sum_jS_B(\alpha_j) \pa_t a_j
\ee
and defines a local function called {\em Bergmann tau function} of the covering.
\end{proposition}
In \cite{KokotovKorotkin1} (and later works by the same authors) explicit expressions for $\tau_{Bx}$ where also derived in terms of Theta-functions. We refer to their work for the explicit expressions since it would lead us too far astray.

We can however conclude by identifying the term $(A)$ in (\ref{eq:w11}) as follows 
\bea
(A) =  \d{x}(p)\frac{\pa}{\pa V(q)} \ln \tau_{Bx}^{-12}
\eea
and hence -finally-
\be
\d{x}(p)\frac{\pa}{\pa V(q)} \F^{(1)} = -\omega^{(1)} =- \frac{1}{24}\d{x}(p)\frac{\pa}{\pa V(q)} \ln \left( \tau_{Bx}^{12} \prod_{j} y'(\alpha_j)\ri)
\ee
One can verify that also the variations with respect to the other moduli (filling fractions, total charge etc.) coincide and hence we can set
\be
\F^{(1)}  = - \frac 1{24} \ln \left( \tau_{Bx}^{12} \prod_{j} y'(\alpha_j)\ri)
\ee

\bibliographystyle{unsrt}
\bibliography{Biblio}

\begin{thebibliography}{10}

\bibitem{BeGeSz-08.2}
M~Bertola, M~Gekhtman, and J~Szmigielski.
\newblock The {C}auchy two--matrix model.
\newblock {\em Comm. Math. Phys (to appear)}, 2009.

\bibitem{EyOr-05.1}
Bertrand Eynard and Nicolas Orantin.
\newblock Topological expansion of the 2-matrix model correlation functions:
  diagrammatic rules for a residue formula functions: diagrammatic rules for a
  residue formula.
\newblock {\em J. High Energy Phys.}, (12):034, 44 pp. (electronic), 2005.

\bibitem{ChEy-06.1}
Leonid Chekhov and Bertrand Eynard.
\newblock Hermitian matrix model free energy: {F}eynman graph technique for all
  genera.
\newblock {\em J. High Energy Phys.}, (3):014, 18 pp. (electronic), 2006.

\bibitem{diFGiZiJ-95.1}
P.~Di~Francesco, Paul~H. Ginsparg, and Jean Zinn-Justin.
\newblock {2-D Gravity and random matrices}.
\newblock {\em Phys. Rept.}, 254:1--133, 1995.

\bibitem{BrItPaZu-78.1}
E.~Brezin, C.~Itzykson, G.~Parisi, and J.B. Zuber.
\newblock {Planar Graphs}.
\newblock {\em Comm. Math. Phys.}, 59:35, 1978.

\bibitem{Da-85.1}
F.~David.
\newblock {Planar diagrams, two-dimensional lattice gravity and surface
  models}.
\newblock {\em Nucl. Phys. B}, 257:45, 1985.

\bibitem{KaKoMi-85.1}
I.K.~Kostov V.A.~Kazakov and A.A. Migdal.
\newblock Critical properties of randomly triangulated planar random surfaces.
\newblock {\em Phys. Lett. B}, 150(Issue 4):282 -- 284, 1985.

\bibitem{Ey-03.1}
Bertrand Eynard.
\newblock Large-{$N$} expansion of the 2-matrix model.
\newblock {\em J. High Energy Phys.}, (1):051, 38, 2003.

\bibitem{ChEyOr-06.1}
Leonid Chekhov, Bertrand Eynard, and Nicolas Orantin.
\newblock Free energy topological expansion for the 2-matrix model.
\newblock {\em J. High Energy Phys.}, (12):053, 31 pp. (electronic), 2006.

\bibitem{Ey-04.1}
Bertrand Eynard.
\newblock Topological expansion for the 1-{H}ermitian matrix model correlation
  functions.
\newblock {\em J. High Energy Phys.}, (11):031, 35 pp. (electronic) (2005),
  2004.

\bibitem{EyPr-08.1}
Bertrand Eynard and Aleix~Prats Ferrer.
\newblock Topological expansion of the chain of matrices, 2008.

\bibitem{EyOr-07.1}
B.~Eynard and N.~Orantin.
\newblock Invariants of algebraic curves and topological expansion.
\newblock {\em Commun. Number Theory Phys.}, 1(2):347--452, 2007.

\bibitem{EyMaOr-07.1}
Bertrand Eynard, Nicolas Orantin, and Marcos Mari{\~n}o.
\newblock Holomorphic anomaly and matrix models.
\newblock {\em J. High Energy Phys.}, (6):058, 20 pp. (electronic), 2007.

\bibitem{EyOr-07.2}
Bertrand Eynard and Nicolas Orantin.
\newblock Weil-{P}etersson volume of moduli spaces, {M}irzakhani's recursion
  and matrix models, 2007.

\bibitem{Ey-07.1}
Bertrand Eynard.
\newblock Recursion between {M}umford volumes of moduli spaces, 2007.

\bibitem{HaOr-06.1}
J.~Harnad and A.~Yu. Orlov.
\newblock Fermionic construction of partition functions for two-matrix models
  and perturbative {S}chur function expansions.
\newblock {\em J. Phys. A}, 39(28):8783--8809, 2006.

\bibitem{BertolaBalogh}
M~Bertola and F~Balogh.
\newblock Regularity of a vector potential problem and its spectral curve.
\newblock {\em Journal of Approximation Theory}, In press, 2009.

\bibitem{Bert-06.1}
M.~Bertola.
\newblock Two-matrix model with semiclassical potentials and extended {W}hitham
  hierarchy.
\newblock {\em J. Phys. A}, 39(28):8823--8855, 2006.

\bibitem{EyKoKo-04.1}
B.~Eynard, A.~Kokotov, and D.~Korotkin.
\newblock {Genus one contribution to free energy in hermitian two- matrix
  model}.
\newblock {\em Nucl. Phys.}, B694:443--472, 2004.

\bibitem{EyKoKo-05.1}
B.~Eynard, A.~Kokotov, and D.~Korotkin.
\newblock {{$1/N^2$} correction to free energy in hermitian two-matrix model}.
\newblock {\em Lett. Math. Phys.}, 71:199--207, 2005.

\bibitem{FarkasKra}
H.~M. Farkas and I.~Kra.
\newblock {\em Riemann surfaces}, volume~71 of {\em Graduate Texts in
  Mathematics}.
\newblock Springer-Verlag, New York, second edition, 1992.

\bibitem{Faybook}
John~D. Fay.
\newblock {\em Theta functions on {R}iemann surfaces}.
\newblock Lecture Notes in Mathematics, Vol. 352. Springer-Verlag, Berlin,
  1973.

\bibitem{KokotovKorotkin1}
A.~Kokotov and D.~Korotkin.
\newblock Tau-functions on {H}urwitz spaces.
\newblock {\em Math. Phys. Anal. Geom.}, 7(1):47--96, 2004.

\end{thebibliography}
\end{document}